\documentclass{article}

\usepackage{chngcntr}
\usepackage{graphicx}
\usepackage{sidecap}
\usepackage{wrapfig}
\usepackage{setspace}
\usepackage{rotating}
\usepackage{amsmath}
\usepackage{endnotes}
\usepackage{hyperref}
\usepackage[left=3cm,top=3cm,right=3cm,bottom=3cm]{geometry}
\usepackage{chngcntr}
\usepackage{soul}
\usepackage{pdflscape}
\usepackage{lineno}
\usepackage{caption}
\usepackage{enumerate}
\usepackage{authblk}
\usepackage{epstopdf}
\usepackage{tabularx}
\usepackage{booktabs}
\usepackage{times}
\usepackage{natbib} %\cite{citeref} \citep{citeref} \citep[e.g.][]{citeref}
\usepackage{bbm}

\usepackage{amssymb}
\usepackage{gensymb}
\usepackage{floatrow}
\usepackage{dcolumn}
\usepackage{rotating}
\usepackage{array} 
\usepackage{changepage}
\usepackage{pdflscape}
\usepackage{enumitem}

\usepackage[toc,page,header]{appendix}
\usepackage{minitoc}
\tightmtctrue 
\noptcrule 
\noptcpagenumbers %partial toc page numbers off

\hypersetup{
    bookmarks=true,         % show bookmarks bar?
    unicode=false,          % non-Latin characters in AcrobatÕs bookmarks
    pdftoolbar=true,        % show AcrobatÕs toolbar?
    pdfmenubar=true,        % show AcrobatÕs menu?
    pdffitwindow=true,     % window fit to page when opened
    pdfstartview={FitH},    % fits the width of the page to the window
    pdftitle={},    % title
    pdfauthor={},     % author
    pdfsubject={},   % subject of the document
    pdfcreator={},   % creator of the document
    pdfkeywords={}, % list of keywords
    pdfnewwindow=true,      % links in new window
    colorlinks=true,       % false: boxed links; true: colored links
    linkcolor=black,          % color of internal links
    citecolor=black,        % color of links to bibliography
    filecolor=black,      % color of file links
    urlcolor=black     % color of external links
}

\newcommand{\beq}{\begin{equation}}
\newcommand{\eeq}{\end{equation}}

\newcommand{\bi}{\begin{itemize}}
\newcommand{\ei}{\end{itemize}}
\newcommand{\mb}{\mathbf}

\newcommand{\bFig}{\begin{figure} \centering}
\newcommand{\eFig}{\end{figure}}
\newcommand{\ben}{\begin{enumerate}}
\newcommand{\een}{\end{enumerate}}

\title{{Accounting for Unobservable Heterogeneity in Cross Section Using Spatial First Differences}\footnote{We thank Wolfram Schlenker for generously sharing data and Michael Anderson, Max Auffhammer, Kendon Bell, Ian Bolliger, Tamma Carleton, Cl\'{e}ment de Chaisemartin, Robert Chirinko, Avi Feller, Andrew Hultgren, Larry Karp, Jeremy Magruder, Aprajit Mahajan, Gordon McCord, Craig McIntosh, Jonathan Proctor, James Stock, and seminar participants at UC Berkeley, UC Santa Barbara, University of Southern California, Stanford University, the University of California Environmental Economics Working Group, and the Association for Environment and Resource Economists for discussions and useful comments.  Druckenmiller is supported by the National Science Foundation Graduate Research Fellowship under Grant No. DGE 1106400. Code is available at \href{http://globalpolicy.science/code}{\texttt{globalpolicy.science/code}}. Emails: \href{mailto:hdruckenmiller@berkeley.edu}{\texttt{hdruckenmiller@berkeley.edu}}, \href{mailto:shsiang@berkeley.edu}{\texttt{shsiang@berkeley.edu}}. }\vspace{5mm}}

\author
{Hannah Druckenmiller$^{1}$ and Solomon Hsiang$^{1,2}$\\
\normalsize{$^{1}$University of California, Berkeley} and 
\normalsize{$^{2}$National Bureau of Economic Research}
}

\begin{document}
\doparttoc % Tell to minitoc to generate a toc for the parts
\faketableofcontents % Run a fake tableofcontents command for the partocs

\maketitle

\onehalfspacing

%---------------------------------------ABSTRACT
\begin{abstract}

\noindent We develop a cross-sectional research design to identify causal effects in the presence of unobservable heterogeneity without instruments. When units are dense in physical space, it may be sufficient to regress the ``spatial first differences'' (SFD) of the outcome on the treatment and omit all covariates. The identifying assumptions of SFD are similar in  mathematical structure and plausibility to other quasi-experimental designs.  We use SFD to obtain new estimates for the effects of time-invariant geographic factors, soil and climate, on long-run agricultural productivities --- relationships crucial for economic decisions, such as land management and climate policy, but notoriously confounded by unobservables.

\end{abstract}

\thispagestyle{empty}

%\doublespacing
\newpage
%\linenumbers

\setcounter{page}{1}

%-------------------------------------INTRO 

\section*{Introduction}

We consider the problem of estimating causal effects in cross-sectional regressions when important covariates, which influence outcomes and are thought to be correlated with the treatment, cannot be observed. It is well understood that the omission of these variables may lead to substantial bias in standard regression approaches to inference.  Here we propose a new cross-sectional research design that is capable of recovering such causal effects even in the presence of omitted variables. We demonstrate the performance of this approach in simulation and in two real data sets by intentionally withholding important covariates during estimation, thereby mimicking contexts with omitted variables. The first application demonstrates the ability of the procedure to recover a well established relationship (returns to schooling) while the second recovers new relationships (geographical determinants of agricultural productivities). The core insight of our approach is that unobserved heterogeneity in many cross-sectional contexts is captured by trends in outcomes across space, which can be understood as a non-parametric component of partially linear semiparametric models \citep{robinson1988root}.  Recognizing this, we suggest that omitted variables bias due to this heterogeneity can be eliminated from estimates using a simple and general differencing approach \citep{yatchew1997elementary} in situations where the spatial position of observations can be located.

When units of observation are organized and densely packed across physical space---such as gridded data or county-level data---we propose an estimator that only compares observations to their immediately adjacent neighbors and simultaneously compares all observations to a neighbor. This approach assumes that immediately adjacent observational units are comparable to one another but does not assume that distant units are comparable, as is assumed in standard cross-sectional approaches. By restricting comparisons to adjacent neighbors in our procedure, the influence of all omitted variables that are common to neighboring units are differenced out. This fundamentally transforms the identifying assumption of cross-sectional analysis into one that matches modern quasi-experimental research designs, such as regression discontinuity designs, in terms of its mathematical structure and plausibility. Conceptually, this approach is similar to using first differences over time in a panel regression to purge data of unobserved factors specific to a panel unit, however in our case the unobserved factors are shared by two observations that are adjacent in space rather than adjacent in time. In fact, our approach is essentially identical, mathematically, to the well known first differences (FD) estimator where the key alteration is to exchange the time index of observations to an index describing the position of observations in space. For this reason, we call our research design ``spatial first differences" (SFD).

%-------------------------------------RESEARCH DESIGN

\section*{The Spatial First Differences Research Design}

In the standard cross-sectional multiple regression research design, we often study situations where an outcome of interest $y$ is influenced by a vector of $K$ observable variables $\mb{x}$ (``treatments'') and might possibly be influenced by $M$ unobservable variables $\mb{c}$ as well:
\beq
y_i = \mb{x}_i\beta + \mb{c}_i\alpha + \epsilon_i 
\label{Eq:XsectionSetup}
\eeq
where $\epsilon$ is an i.i.d. disturbance term with mean zero. $N$ observational units are indexed by $i$. The $K$ parameters of interest are estimates of the causal effects in the vector $\mb{\beta}$. It is well known that if $\mb{c}_i$ is omitted from the cross-sectional ordinary least-squares (OLS) regression in levels (denoted by subscript $L$)
\beq
y_i = \mb{x}_i\hat{\beta}_{L}  + \hat{\epsilon}_i 
\label{Eq:XsectionOLS}
\eeq
then the ``omitted variables bias" in the vector of parameter estimates is
$$
E[\hat{\beta}_{L}-\beta] = E[(\mb{x}'\mb{x})^{-1}(\mb{x}'\mb{c}\alpha)]
$$
which may be large if the covariance between $\mb{x}_i$ and $\mb{c}_i$ is large and/or if any of the $M$ elements in $\alpha$ are large. However, because $\mb{c}$ is unobserved, it is not generally possible to know whether either of these conditions apply. Due to this fact, the specter of omitted variables bias now looms large over cross-sectional regression analyses and $\hat{\beta}_{L}$ is often assumed to be biased unless corroborated using an alternative research design. Thus, in many fields, $\hat{\beta}_L$ is no longer used as a basis for causal inference  \citep[e.g.][]{leamer1983let,holland1986statistics,clarke2005phantom, angrist2010credibility}, regardless of how many covariates are included in the regression model.

The weakness of the standard cross-sectional research design results from how it addresses the fundamental challenge of causal inference, i.e. the estimation of plausible counterfactual outcomes \citep{holland1986statistics}. For a change of $\mb{x}$ from $\mb{x}_j$ to $\mb{x}_i$, the average treatment effect of interest from Eq. (\ref{Eq:XsectionSetup}) is 
\beq
(\mb{x}_i-\mb{x}_j)\beta = E[y_i|\mb{x}_i]-E[y_i|\mb{x}_j]
\label{Eq:XsectionBeta}
\eeq
where $E[y_i|\mb{x}_j]$ is the expected potential outcome for observational unit $i$ if it were treated with the $\mb{x}$'s of observation $j$. However, since this term is never observed in the real world, a researcher estimating Eq. (\ref{Eq:XsectionOLS}) assumes
\beq
E[y_i|\mb{x}_j]=E[y_j|\mb{x}_j]\;\;\forall \;\; i\neq j
\label{Eq:StrongUnitHomogeniety}
\eeq
which states that the expected potential outcome for $i$ and the outcome for $j$ would be the same if both units were treated with $\mb{x}_j$, which in reality was only received by $j$ and not $i$. This Conditional Independence Assumption is a relatively strong assumption in many contexts because it assumes \textit{all} observational units in a cross section of data are comparable. Substituting Eq. (\ref{Eq:StrongUnitHomogeniety}) into Eq. (\ref{Eq:XsectionBeta}) delivers the standard cross-sectional research design in levels, which provides an unbiased estimate of treatment effects if Eq. (\ref{Eq:StrongUnitHomogeniety}) is true. However, in the presence of unobserved heterogeneity, such as the variables described by $\mb{c}$ in Eq. (\ref{Eq:XsectionSetup}), then the assumption in Eq. (\ref{Eq:StrongUnitHomogeniety}) will not be true since units $i$ and $j$ will not longer be comparable when conditioned only on their $\mb{x}$'s. 

We propose that the treatment effect in Eq. (\ref{Eq:XsectionBeta}) can sometimes be credibly identified in the presence of unobserved $\mb{c}$'s by reformulating the estimation procedure to only compare small differences in $\mb{x}$ and $y$ between adjacent observational units. This approach exploits a conditional independance assumption that is dramatically weaker than Eq. (\ref{Eq:StrongUnitHomogeniety}) because observational units are only compared to their immediately adjacent neighbors. If $i$ is an index that matches the rank-order of observations across space in an arbitrary coordinate system, such that observations $i$ and $i-1$ are immediately adjacent to one another, then the SFD research design replaces the assumption of Eq. (\ref{Eq:StrongUnitHomogeniety}) with the substantially weaker assumption
\beq
E[y_i|\mb{x}_{i-1}]=E[y_{i-1}|\mb{x}_{i-1}] \;\; \forall \;\; \{i,i-1\}
\label{Eq:LocalUnitHomogeniety}
\eeq
which states that the expected potential outcome for two immediate neighbors $i$ and $i-1$ are equal if they were to receive the same treatment $\mb{x}_{i-1}$. Eq. (\ref{Eq:LocalUnitHomogeniety}) is a strictly weaker assumption than Eq. (\ref{Eq:StrongUnitHomogeniety}) because the latter holds for \textit{all} pairs of observations in the sample, whereas Eq. (\ref{Eq:LocalUnitHomogeniety}) states that the same conditions hold only for the subset of pairs where the observations are adjacent.\footnote{It may be tempting to suggest that Eq. (\ref{Eq:LocalUnitHomogeniety}) necessarily implies Eq. (\ref{Eq:StrongUnitHomogeniety}) because of transitivity, but this is not the case. To see why, note that $E[y_i|\mb{x}_{i-1}]=E[y_{i-1}|\mb{x}_{i-1}]$ and $E[y_{i+1}|\mb{x}_{i}]=E[y_{i}|\mb{x}_{i}]$ do not share common terms, since the former is conditioned on $\mb{x}_{i-1}$ and the latter on $\mb{x}_{i}$.} Because Eq. (\ref{Eq:LocalUnitHomogeniety}) imposes that units are only conditionally independent with respect to their local neighbors, we denote it the \textit{Local Conditional Independence Assumption}. 

Conditions under which Eq. (\ref{Eq:LocalUnitHomogeniety}) is plausible are discussed below, but it is worth noting at the outset that this assumption is conceptually similar to (\textit{i}) the assumption that immediately sequential observations within a time-series are comparable
\beq
E[y_{t}|\mb{x}_{t-1}]=E[y_{t-1}|\mb{x}_{t-1}] \;\; \forall \;\; \{t,t-1\},
\label{Eq:TSFD}
\eeq
the assumption exploited in event-study research designs and many FD time-series models; (\textit{ii}) the assumption that sequential observations within a panel unit are comparable
$$
E[y_{it}|\mb{x}_{i,t-1}]=E[y_{i,t-1}|\mb{x}_{i,t-1}]  \;\; \forall \;\; \{t,t-1 \mid i\},
$$
an assumption exploited in differences-in-differences panel research designs\footnote{These are a subset of the assumptions required for many differences-in-differences research designs, since these approaches also often assume common trends across panel units.} (e.g. panel fixed-effects estimators); and (\textit{iii}) the assumption that observations just above and just below a treatment discontinuity are comparable
\beq
E[y_{above}|\mb{x}_{below}]=E[y_{below}|\mb{x}_{below}],
\label{Eq:RD}
\eeq
the assumption exploited in regression discontinuity research designs. In fact, the SFD research design is, mathematically speaking, almost identical to the FD approach in times-series and panel analysis \citep[e.g.][]{wooldridge2010econometric}, except the one-for-one transposition of time and space indices---a similarity that allows researchers to easily implement SFD by ``tricking'' software packages into using time series operators on cross-sectional data sets by substituting the spatial indices for time indices. We also note that, if it is helpful, one may also think of the SFD research design ``{as if}'' the researcher is simultaneously running a large number of regression discontinuity research designs in space, in the sense of \cite{black1999better}, with one ``discontinuity'' for every pair of adjacent observations in the SFD setup. Thus, overall, we argue that the Local Conditional Independence Assumption required by the SFD research design is at least as valid as corresponding assumptions in other widely accepted identification strategies, when each is applied to the appropriate context.

\begin{figure}
\begin{center}
\includegraphics[width=\textwidth]{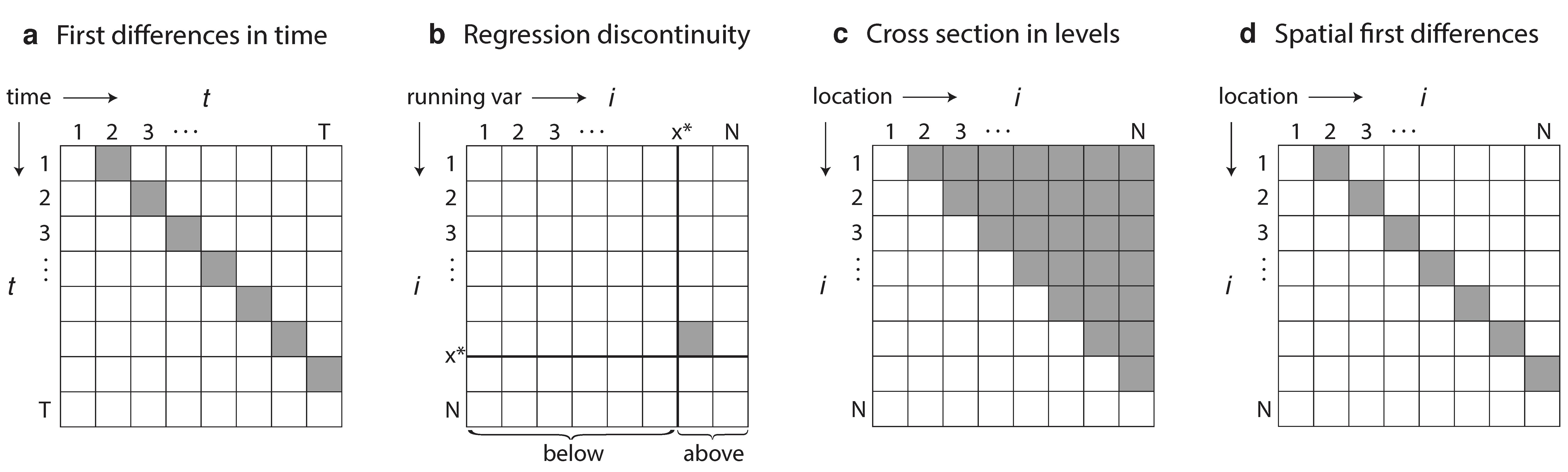}
\caption{{\bf Comparison of pair-wise assumptions regarding the comparability of observations needed for identification in different research designs.} Graphical depiction of the various comparisons exploited to identify causal effects in (a) FD time-series models, (b) regression discontinuity designs with discontinuity at $x^*$, (c) the cross-sectional approach in levels, and (d) SFD. Each observation in a data set appears on both a row and column for a grid. Squares are grey if the observations for that row and column are assumed to be comparable (i.e. expected potential outcomes are conditionally equal) when using the associated research design.}
\label{Fig:AssumptionMatrix}
\end{center}
\end{figure}

To illustrate how SFD brings the number of identifying assumptions needed for cross-sectional analyses into parity with the research designs described above, Figure \ref{Fig:AssumptionMatrix} graphically depicts the comparisons exploited to identify causal effects in different research designs. Each grid depicts a different research design, and each observation in a data set appears on both a row and column of that grid. A square is shaded grey if the two observations corresponding to that row and column are assumed to be comparable when using the associated research design (pairs are only shaded once). Panel a illustrates how only observations that are adjacent in time are compared to one another in FD time-series models (Eq. \ref{Eq:TSFD}).  Panel b displays how only observations with running variable values just below and just above the cutoff value $x^*$ are assumed to be comparable in a regression discontinuity design (Eq. \ref{Eq:RD}). Panel c shows the large number of $(N-1)\frac{N}{2}$ comparisons made when using the standard ``levels'' approach to cross-sectional research designs (Eq. \ref{Eq:StrongUnitHomogeniety}), where every observation is compared to every other observation. Panel d demonstrates how SFD reduces the number of comparisons to a strict subset of the comparisons in the levels model, since observations are only compared to those immediately adjacent in space (Eq. \ref{Eq:LocalUnitHomogeniety}). The necessary $N-1$ assumptions in SFD regarding the comparability of neighbors (panel d) resembles the $T-1$ assumptions in FD time-series models (panel a); or, alternatively, $N-1$ different regression discontinuity analyses (panel b) executed across space. In contrast, the strong Conditional Independence Assumption necessary for the cross-sectional levels approach (panel c) requires exactly $\frac{N}{2}$ \textit{times} as many pair-wise assumptions (compared to SFD) in order to identify $\hat{\beta}$.

\subsection*{Estimation of Spatial First Differences}

We exploit the Local Conditional Independence Assumption by comparing each observation to a spatially adjacent neighbor. Critically, we difference each pair of neighboring observations to purge the data of unobserved factors that are common to the pair. We thus construct the SFD approach by writing Eq. (\ref{Eq:XsectionSetup}) for spatially adjacent observations $i$ and $i-1$, and then differencing the equations
\beq
\underbrace{y_{i}-y_{i-1}}_{\Delta y_{i}} = \underbrace{(\mb{x}_{i}-\mb{x}_{i-1})}_{\Delta \mb{x}_i}\beta + \underbrace{(\mb{c}_{i}-\mb{c}_{i-1})}_{\Delta\mb{c}_i}\alpha + \underbrace{(\epsilon_{i}-\epsilon_{i-1})}_{\Delta \epsilon_i}
\label{Eq:SFDTheory}
\eeq 
where the $\Delta$ operator is analogous to the difference operator in time series analysis. Here we use the convention of denoting differences by the index of the higher-valued index of the pair of observations ($i$ rather than $i-1$). Because $\Delta\mb{c}_i$ cannot be observed, it does not appear in the SFD regression model 
\beq
\Delta y_i = \Delta \mb{x}_i \beta_{SFD} + {\Delta {\mb{\epsilon}}}_i
\label{Eq:SFDModel}
\eeq
which can be estimated using any regression solution, such as maximum likelihood or ridge regression. We focus here on OLS, since it is straightforward to solve and its properties are widely understood.  The OLS estimator for an SFD research design is
\beq
\hat{\beta}_{SFD} = (\Delta \mb{x}'\Delta \mb{x})^{-1}(\Delta \mb{x}'\Delta y).
\label{Eq:SFDEstimator}
\eeq
where as usual a vector of ones is included in the $\Delta \mb{x}$ matrix to guarantee variables are centered and errors are mean zero, although this is not explicit in our notation for parsimony.\footnote{This ``constant'' term is a nuisance parameter, discussed below.} 

If the local conditional independence assumption Eq. (\ref{Eq:LocalUnitHomogeniety}) holds, then it implies
\beq
E[\Delta\mb{x}'\Delta\mb{c}]=0_{K,M}
\label{Eq:IndepCondition}
\eeq
where $0_{K,M}$ is the $K\times M$ null matrix. Eq. (\ref{Eq:IndepCondition}) is the core identifying assumption of SFD when it is solved by OLS, stating that the covariance between \textit{changes in} $\mb{x}$ and \textit{changes in} $\mb{c}$ between immediately adjacent neighbors is not systematically correlated within the population.\footnote{See the Appendix for an alternative interpretation of Eq. (\ref{Eq:IndepCondition}) based on isotropy.} The plausibility of identification via SFD depends on the validity of Eq. (\ref{Eq:IndepCondition}), which is actually weaker than the local conditional independence assumption; although we emphasize the latter above because it is a more easily conceptualized heuristic for considering the validity of SFD in applied settings and, if true, it implies Eq (\ref{Eq:IndepCondition}) holds.

Note that the orthogonality condition in Eq. (\ref{Eq:IndepCondition}) describes the covariance between the local derivatives of $\mb{x}$ and $\mb{c}$ with respect to space. Thus violation of Eq. (\ref{Eq:IndepCondition}) only occurs if the second derivatives of these variables with respect to space are sufficiently synchronized that they systematically induce changes in these derivatives in the same locations.  The assumption in Eq. (\ref{Eq:IndepCondition}) therefore differs \textit{fundamentally} from the identifying assumption in the levels model (Eq. \ref{Eq:XsectionOLS}):
\beq
E[\mb{x}'\mb{c}]=0_{K,M}
\label{Eq:LevelsIdentification}
\eeq
which describes covariance between the levels of $\mb{x}$ and $\mb{c}$ across the entire sample.  In general, the failure of Eq. (\ref{Eq:LevelsIdentification}) implies nothing about the validity of Eq. (\ref{Eq:IndepCondition}), thus a levels-based cross-sectional approach may be invalid at the same time that the analogous SFD cross-sectional approach is valid. We discuss this point further below.
 
If the identifying assumption in Eq (\ref{Eq:IndepCondition}) is true, then we have
\beq
E[\hat{\beta}_{SFD}-\beta] = E[(\Delta\mb{x}'\Delta\mb{x})^{-1}(\Delta\mb{x}'\Delta\mb{c}\alpha)] = 0_{K,1}
\label{Eq:SFDBias}
\eeq
so $\hat{\beta}_{SFD}$ will be unbiased.  Intuitively, if $\mb{c}$ is common between neighbors, its influence on $y$ will be differenced out and $\Delta \mb{c}$ will be very near or equal zero. Should there exists a component of $\mb{c}$ that is not common between neighbors, $\hat{\beta}_{SFD}$ will still be unbiased so long as the non-zero component of $\Delta \mb{c}$ is uncorrelated with changes in $\mb{x}$ between neighbors ($\Delta \mb{x}$). Thus $\hat{\beta}_{SFD}$ is generally robust to unobservable to heterogeneity in factors that are spatially correlated ($\mb{c}_i\approx\mb{c}_{i-1}$) and factors that are i.i.d with respect to spatial position (Eq. \ref{Eq:IndepCondition}). As we demonstrate in later sections, the SFD approach effectively purges estimates of the influence of unobserved factors $\mb{c}$ across a variety of applied contexts.

\subsection*{Asymptotic distribution and estimation of variance}

As described here, $\hat{\beta}_{SFD}$ falls within a class of difference estimators explored by \cite{yatchew1997elementary, yatchew1999differencing}.  To our knowledge, Yatchew did not discuss applying those results to an explicitly spatial context to identify causal effects---nonetheless, Yatchew's results apply here since our context is a specific case of that more general problem. \cite{yatchew1997elementary} demonstrated that under mild conditions, estimation of $\hat{\beta}_{SFD}$ via OLS is consistent, yielding an asymptotic distribution 
\beq
\hat{\beta}_{SFD} \mathop{\sim}\limits^{A} \mathcal{N}\left(\beta,\; \frac{1.5 \sigma^2_{\epsilon}}{N}\Omega^{-1}_{x}\right) \label{Eq:Asymp}
\eeq
as the number of observations increases to infinity and the spatial distance between observations vanishes. Here $\sigma^2_{\epsilon}$ is the population variance of $\epsilon$ and $\Omega_{x}={E}[Cov(\mb{x}|\ell_i)]$ where $\ell_i$ is the spatial position of observation $i$. \cite{yatchew1997elementary} also demonstrated that these variances can be estimated consistently:
\begin{align}
\hat{s}^2_{\epsilon, SFD}= & \frac{1}{N}\sum_{i=2}^N \hat{\Delta \epsilon}_i^2= \frac{1}{N}\sum_{i=2}^N (\Delta y_i - \Delta \mb{x}_i \hat{\beta}_{SFD})^2\mathop{\rightarrow}\limits^{P}\sigma_\epsilon^2 \label{Eq:s2} \\
\hat{\Omega}_{x,SFD} = & \frac{1}{N}\Delta \mb{x}'\Delta \mb{x}\mathop{\rightarrow}\limits^{P}\Omega_x.
\end{align}
As can be seen in Eq. (\ref{Eq:Asymp}), $\hat{\beta}_{SFD}$ does not achieve the Cramer-Rao lower bound\footnote{\cite{yatchew1997elementary} suggests an optimal combination of higher-order differencing estimates to achieve this bound asymptotically, a result that should in theory apply to the SFD context, but whose practical exploration we leave to future work.} and instead has an efficiency of 66.7\% relative to this bound. In many applied context where ample data is available, we think this sacrifice of efficiency may be reasonable in order to obtain an unbiased cross-sectional estimate. 

Importantly, in finite samples, the usual OLS estimator $\hat{s}^2_{\epsilon}$ in Eq. (\ref{Eq:s2}) is not appropriate because ${\Delta {\mb{\epsilon}}}$ will be autocorrelated between adjacent units (eg. $\hat{\Delta {\epsilon}}_i$ and ${\hat{\Delta {\epsilon}}}_{i+1}$ both contain ${\epsilon}_i$). Thus, in practice, we recommend that the variance of  $\hat{\beta}_{SFD}$ be estimated using the autocorrelation-robust approaches described by \cite{newey87} and \cite{conley1999gmm}, allowing for autocorrelation in disturbances between nearby observations (in one and two-dimensions, respectively) after differencing.\footnote{Differencing requires the use of a kernel that spans at least one adjacent unit in each direction when estimating the covariance matrix for $\hat{\Delta \epsilon}$, but larger kernels may be appropriate if $\hat{\Delta {\epsilon}}$ is spatially correlated across larger scales, as in the maize example we consider below.} In the Appendix, we show that this approach generally provides the most conservative inferences relative to alternatives in our empirical application.

\subsection*{Implementation in a one-dimensional space}

\begin{figure}[t]
\begin{center}
\includegraphics[width=\textwidth]{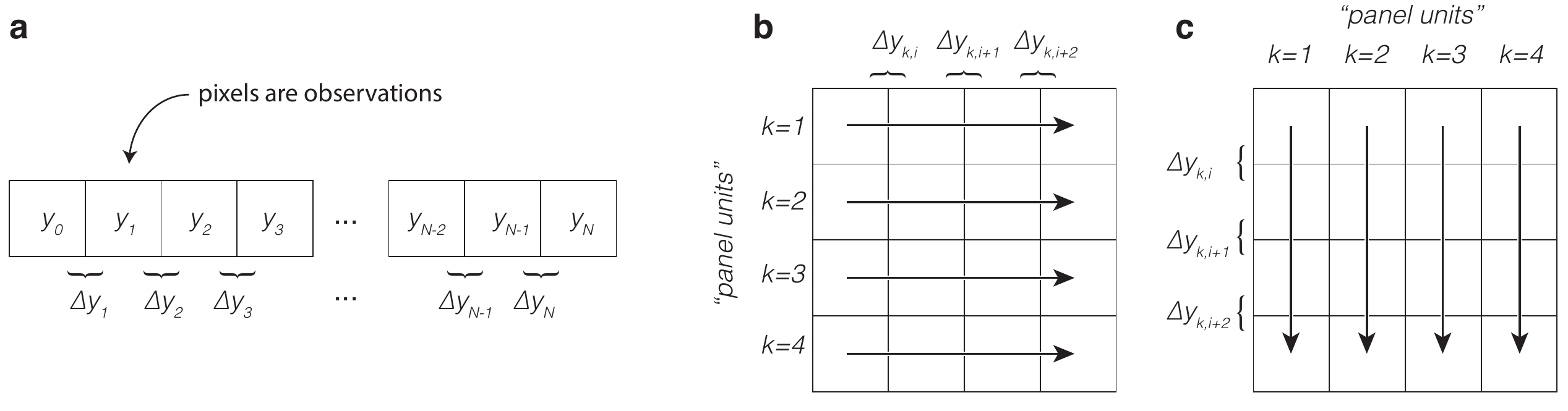}
\caption{{\bf Implementation of spatial first differences using gridded data.} Each square represents the location of an observation. (a) Implementation of SFD in one-dimensional space using a regular grid. The $i$th observation in the differenced data set contains the change in both the treatment $(\Delta \mb{x}_i)$ and the outcome $(\Delta {y}_i)$ between immediately adjacent neighbors in positions $i$ and $i-1$. Only $\Delta {y}_i$ is shown. (b) Implementation of SFD in two-dimensional gridded data, where differences are computed in the West-East direction. Each row of observations (here indexed by $k$) is analogous to a panel unit in panel data. (c) Same, but differences are computed in the North-South direction.}
\label{Fig:Implementation}
\end{center}
\end{figure}

In the simplest case, the physical space in which observations are located has only one dimension, such as households located along a road. Panel a in Figure \ref{Fig:Implementation} depicts this setup, where the $i$th observation in the differenced data set contains the change in both the treatment ($\Delta \mb{x}_i$) and the outcome ($\Delta y_i$) between immediately adjacent neighbors in positions $i$ and $i-1$. In this situation, the setup is directly analogous to FD in time series analysis, where the only change is that the position of an observation in time is replaced by the position of an observation along the one-dimensional space. When data is arranged in this way, it is straightforward to estimate SFD by applying basic time series functions that are standard in most statistical packages. For example, a researcher might estimate the effect of \texttt{years\_of\_education} on \texttt{wages} among individuals living at position \texttt{house\_number} along a single road. Implementing Eq. (\ref{Eq:SFDModel}) via OLS in the statistical package Stata would then only require the two commands \footnote{In the statistical package R, the same procedure is implemented by the two commands: \\

\indent $\quad \quad$ \texttt{dplyr::arrange(data, house\_number)} \\
\indent $\quad \quad$ \texttt{lm(diff(wages) $\sim$ diff(years\_of\_education), data)} \\

and in Python, one could implement this procedure using pandas after importing statsmodels.formula.api as sm: \\

\indent $\quad \quad$ \texttt{data.sort\_values(by=[`house\_number'])} \\
\indent $\quad \quad$ \texttt{diff\_data = data.diff()} \\
\indent $\quad \quad$ \texttt{model = sm.ols(formula= "wages $\sim$ years\_of\_education", data=diff\_data).fit()}} \\

\indent \texttt{tsset house\_number}\\
\indent \texttt{regress D.wages D.years\_of\_education}\\

\noindent where the first command ``tricks'' the software by telling it that the data is a ``time series'' where the time variable is \texttt{house\_number}. The second command then exploits the difference operator \texttt{D} which computes first differences in both variables along the road and estimates the SFD regression. Whether the software is informed that ``time'' moves forward as one travels up or down the \texttt{house\_number} variable is irrelevant, the SFD estimate will be the same. 

\subsection*{Implementation in a two-dimensional gridded space}

Implementing SFD in two-dimensional space is similarly simple if data are ``gridded'' on a regular lattice, such as pixels describing topographical ruggedness or night lights. Gridded data sets of this sort are rapidly growing in availability \citep[e.g.][]{donaldson2016}, making this a particularly useful case to consider. Panel b and c in Figure \ref{Fig:Implementation} depict two ways that SFD can be implemented using such gridded data: differences can be computed between neighbors defined in the East-West sense (panel b) or in the North-South sense (panel c). Neighbors that are adjacent along the dimension that is not differenced (e.g. North-South neighbors in panel b) are simply not compared. In this case, SFD is implemented in a manner analogous to FD applied to panel data, where the row (panel b) or column (panel c) of each sequence of differenced observations (indexed by $k$ in Figure \ref{Fig:Implementation}) are analogous to the panel units in the FD model. A researcher interested in the effect of \texttt{ruggedness} on  \texttt{night\_lights} in a gridded data set where pixels are indexed by \texttt{latitude} and \texttt{longitude} could implement the East-West SFD model in \texttt{Stata} using the two commands\\

\indent \texttt{xtset latitude longitude}\\
\indent \texttt{regress D.night\_lights D.ruggedness}\\

\noindent where the first command tells the software to treat \texttt{latitude} as if it were the panel variable and \texttt{longitude} as if it were the time variable in a normal panel dataset. The North-South SFD model could be similarly estimated, but switching which dimension is declared analogous to time\\

\indent \texttt{xtset longitude latitude}\\

\noindent prior to estimating Eq. (\ref{Eq:SFDModel}).  

The ability to estimate $\hat{\beta}_{SFD}$ twice along two orthogonal dimensions of a gridded data set---exploiting entirely different variation in the independent variables---provides a natural and appealing check on the robustness and validity of the two estimates since spatial patterns in omitted variables along one dimension might be different than along the other dimension. 

Implementing SFD in two dimensional data when the data are gridded is straightforward, although it is somewhat more difficult to implement on data sets with irregular spatial structure. Below we demonstrate one approach that produces similar ``panel-like'' data structures (similar to  Figure \ref{Fig:Implementation}b) for the cross section of US counties. Although we first attempt to develop the readers intuition for why the procedure works, provide practical guidance, and consider the performance of SFD under simpler one-dimensional scenarios.

\subsection*{Why it works: elimination of spatially correlated unobserved heterogeneity}

A central benefit of the SFD approach is that it eliminates bias due to all spatially correlated unobserved variables, which in many cross-sectional contexts represents most or all of the important omitted factors $\mb{c}$. For example, in a cross-sectional regression of earnings on years of schooling, if households that have high levels of education tend to live in areas with more Whites and Whites tend to earn more than other races, then race will be a spatially correlated omitted variable if it is not included in the model. SFD eliminates spatially correlated unobserved heterogeneity at two levels: the procedure filters out the influence of all factors that vary at low spatial frequencies (any factor that affects observations that are not immediately adjacent) and it differences out all common influences that idiosyncratically affect any two observations that are immediately adjacent to one another. 

\begin{figure}
\begin{center}
\includegraphics[width=\textwidth]{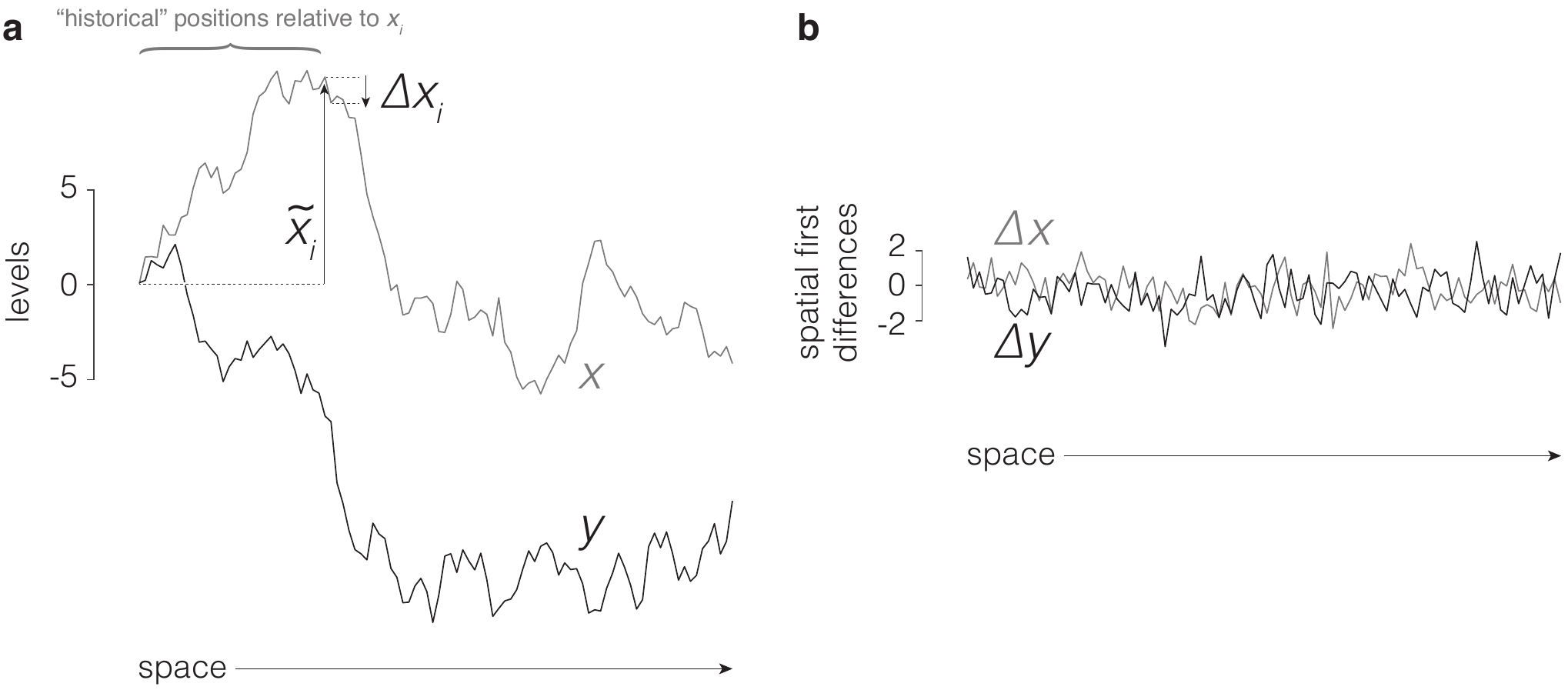}
\caption{{\bf Comparison of levels and spatial first differences.} (a) Levels of $x$ and $y$ as a function of space. (b) SFD in $x$ and $y$ across space. The variables $x$ and $y$ are a unit root with changes $\Delta x$ and $\Delta y$ both distributed $N(0,1)$. The ``spatial history'' is $\tilde{x}_i=x_{i-1}$.}
\label{Fig:SpatialHistories}
\end{center}
\end{figure}

We think a useful way to see the benefit of SFD's high-pass filtering is to consider how $\mb{x}$ and $\mb{c}$ vary as an observer traverses the physical space in the sample, leveraging intuition and language from thinking about cross sections spanning space as analogous to time-series spanning time. Let us  define some arbitrary initial position as $i=1$ (analogous to $t=0$ in time-series) and observe how $\mb{x}$ and $\mb{c}$ evolve as we move sequentially from adjacent neighbor to neighbor away from $i=1$ (analogous to moving forward in time). Then we see that $\mb{x}$ and $\mb{c}$ evolve in a ``unit-root-like'' manner across space because each variable is equal to the sum of its ``spatial history''---i.e. all evolutions of the variables that have occurred since the initial position---and the change in the variable that occurred between the immediately previous position and the current position. Call $\tilde{\mb{x}}_i$ the spatial history of $\mb{x}_i$ where
\beq
\mb{x}_i = \mb{\tilde{x}}_i + \Delta \mb{x}_i
\label{Eq:SpatialHist}
\eeq
and $\mb{\tilde{x}}_i$ represents the accumulation of changes since the arbitrarily defined starting point
$$
\tilde{\mb{x}}_i = \sum_{s=1}^{i-1}\Delta \mb{x}_s.
$$
Define the spatial histories $\tilde{\mb{c}}_i$ and $\tilde{y}_i$ analogously.  These terms are the cumulative effect of all changes in each variable as a path from position $1$ to $i-1$ is traced out through space, analogous to a line-integral of first differences through space.  Panel a of Figure \ref{Fig:SpatialHistories} illustrates this decomposition.

Taking the standard cross-sectional regression shown in Eq. (\ref{Eq:XsectionOLS}), which we hereafter refer to as the ``levels'' model, we know the OLS estimate for $\beta$ is
$$
\hat{\beta}_{L} = \beta + \underbrace{(\mb{x}'\mb{x})^{-1}\mb{x}'(\mb{c}\alpha+\epsilon)}_{\mathrm{bias\;in\;levels\;model}}
$$
where the key term that generates omitted variables bias is $\mb{x}'\mb{c}\alpha$.  The bias originating from this term can be decomposed into contributions from spatial histories and spatial first differences
\begin{align}
\mb{x}'\mb{c}\alpha& = (\mb{\tilde{x}}' + \Delta \mb{x}')(\mb{\tilde{c}}+ \Delta\mb{c})\alpha\notag\\
& = (\underbrace{\mb{\tilde{x}}'\mb{\tilde{c}}}_{\mb{W}_1} + \underbrace{\Delta \mb{x}'\mb{\tilde{c}} +  \mb{\tilde{x}}'\Delta\mb{{c}}}_{\mb{W}_2}+ \underbrace{\Delta \mb{x}'\Delta\mb{c}}_{\mb{W}_3})\alpha
\end{align}
where the total bias depends on the size of elements in the matrices $\mb{W}_1$, $\mb{W}_2$ and $\mb{W}_3$, each of which is $K\times M$. $\mb{W}_1$ is the sample covariance between the spatial histories of $\mb{x}$ and all omitted factors, $\mb{W}_3$ is the sample covariance between their spatial first differences, and $\mb{W}_2$ is the sum of the cross-covariances. 

In most contexts, the most important source of bias is $\mb{W}_1$, which is the sample covariance between the spatial histories of observable and unobservable factors.  This term may be quite large, since any realizations in $\Delta\mb{x}_j$ and $\Delta\mb{c}_l$ that occur within the spatial history of observation $i$ (i.e. $j<i$ and $l<i$) and are correlated will induce correlation in $y_i$ and $\mb{x}_i$. Because each realization of $\Delta\mb{x}$ and $\Delta\mb{c}$ affect \textit{all} observations that occur in their ``spatial future,'' these effects accumulate, causing correlations between $\mb{x}$ and $\mb{c}$ to sometimes grow large in finite samples, even if there is no causal or otherwise mechanical relationship between the two variables. Such correlation is clear in Panel a of Figure \ref{Fig:SpatialHistories}, where the accumulation of i.i.d. realizations (Panel b) generate large correlations between $x$ and $y$ across space that have no causal meaning. In a large number of cross-sectional contexts, the bulk of correlation between $\mb{x}$ and $\mb{c}$ is captured by their spatial histories $\mb{\tilde{x}}$ and $\mb{\tilde{c}}$. In contrast, when the SFD estimator is used, spatial histories and their associated bias are eliminated by differencing (since $\mb{\tilde{x}}_{i}=\mb{x}_{i-1}$), thus all omitted variables biases attributable to $\mb{W}_1$ are purged. Biases due to $\mb{W}_2$ are also purged in the SFD approach, although this is usually only a modest improvement since this term is likely small in most contexts.  Thus the magnitude of of the omitted variables bias that is eliminated from the cross-sectional regression by differencing out spatial histories is
\beq
\hat{\beta}_{L} - \hat{\beta}_{SFD} =(\mb{x}'\mb{x})^{-1}(\mb{W}_1+\mb{W}_2)\alpha
\label{Eq:OVBComparison}
\eeq
which we show below may be substantial. If the condition in Eq. (\ref{Eq:IndepCondition}) is satisfied, which is true if the local conditional independance assumption is valid, then $E[\mb{W}_3]=0$ and $\hat{\beta}_{SFD}$ is identified even if $\hat{\beta}_{L}$ is not. Note that the identifying assumption in Eq. (\ref{Eq:IndepCondition}) does not constrain the magnitude of $\mb{W}_1$, so the validity of the SFD estimator provides no support for the validity of the analogous levels estimator.

The low-frequency spatial correlations between $\mb{x}$ and $\mb{c}$ is a major source of omitted variables bias in many cross-sectional settings, however  spatial correlations between $\mb{x}$ and $\mb{c}$ with high spatial frequencies (analogous to local ``shocks'') may also be problematic in the traditional levels model. For example, if a single unobserved hospital is particularly good at providing healthcare, then average health for individuals in adjacent neighborhoods may be idiosyncratically high, possibly confounding regressions of health on $\mb{x}$. Because the SFD estimator only exploits changes in $\mb{x}$ and $y$ that occur between neighboring observations, any localized change in $\mb{c}$ that affects both locations $i$ and $i-1$ (e.g. the presence of a hospital) is differenced out when $\Delta y_i$ is constructed. 

Thus, in the SFD model, there remains no spatial correlations in unobservables left to bias the estimated parameters. Spatial correlations that affect observations more than one unit away from one another (e.g. $i$ and $i-2$) are purged because the spatial histories of observations are eliminated from the data (observation $i$ does not ``know'' about the existence of observation $i-2$) and spatial correlations that affect adjacent observations (e.g. $i$ and $i-1$) are differenced out.

\subsection*{Relationship of SFD to other models}

SFD is a distinct research design that differs from, but sometimes links to, other approaches that also consider spatial relationships.  Conceptually, the assumptions of SFD generalize the assumptions used in spatial regression discontinuity (RD) \citep[e.g.][]{black1999better} so that obtaining arbitrary boundaries is not a requirement and variation throughout a sample can be exploited. SFD uses spatial relationships strictly to organize observations for the purpose of identification and does not rely on the tools used to handle spatial dependence or spatial autocorrelation developed elsewhere \citep[e.g.][]{anselin1988spatial}. However, we recommend estimating SFD standard errors using the approach described by \cite{newey87} or \cite{conley1999gmm}, in one and two-dimensional contexts, respectively, to account for the possibility that $\hat{\Delta \epsilon}$ is spatially autocorrelated. Additionally, in the presence of spatial spillovers, it is important that spatial lags are included in the regression model (as additional elements in the vector $\mb{x}$) before spatial first differences are computed. At a high level, SFD can be thought of as a simple and general approach to identifying partially linear semiparametric models \citep[e.g.][]{robinson1988root} with unknown omitted variables in contexts where observations are dense and organized across space.  We describe the connections of SFD to these different models in greater detail below.

\paragraph{Spatial regression discontinuity research designs}
Spatial RD designs, following \cite{holmes1998effect}, \cite{black1999better} and others, exploit arbitrary borders in space (e.g. state borders or school districts) and are now widely used in applied research for causal inference. Spatial RD designs rely on the assumption that observations on immediately opposing sides of an arbitrary spatial boundary are conditionally comparable to one another in factors that are not determined by the boundary, an assumption that is the same as the Local Conditional Independence Assumption (Eq. \ref{Eq:LocalUnitHomogeniety}) required by SFD, albeit restricted to apply only to observations near the exploited border. In spatial RD designs, this assumption is often stated as the requirement ``that all confounding factors vary smoothly `enough' in space across the boundary.'' This assumption is important because it implies that changes in treatment $\mb{x}$ across the boundary are not correlated with changes in confounders $\mb{c}$, a statement which written mathematically is precisely the orthogonality condition in Eq. (\ref{Eq:IndepCondition}). 

Thus the underlying assumptions of SFD can be interpreted as a generalization of the assumptions in spatial RD, since in SFD the comparability of immediately adjacent neighbors---only explicitly assumed to apply near a the boundary in spatial RD---is assumed to hold everywhere. Intuitively, if one accepts the spatial RD approach, then extending the assumed local comparability of adjacent observations to units far from the boundary may be natural, since it is often the case that adjacent units far from a boundary (e.g. adjacent counties inside the same state) are at least as comparable to one another as adjacent units on opposite sides of a boundary (e.g. adjacent counties in different states).  Thus, SFD can be thought of ``as if'' it generalizes spatial RD, where small ``discontinuities'' between all pairs of neighbors throughout the sample (not only at borders) are exploited; although the correspondence is more true conceptually than it is exact mathematically, since spatial RD is implemented in a variety of creative ways that do not all map exactly to SFD.\footnote{The generalization of spatial RD using SFD can be seen most easily in one-dimensional space. The univariate spatial RD approach can be rewritten as the SFD estimate, where the treatment $\mb{x}=1$ if observations are on one side of a border and $\mb{x}=0$ on the other side of the border. However, not all implementations of spatial RD are so simple. For example, in some cases spatial RD is implemented by discarding observations far from the boundary, while in other cases these observations are kept and the outcome is allowed to depend flexibly on the running variable (space) far from the boundary, perhaps using splines or nonparametric approaches.} 

\paragraph{Models of spatial dependence}
SFD is fundamentally distinct from the class of methods that collectively compose the field of ``spatial econometrics'' in which regression models are specifically structured to account for relationships that manifest over space across different observational units\footnote{\cite{anselin1988spatial} explicitly defines spatial econometrics as ``the collection of techniques that deal with the peculiarities caused by space in the statistical analysis of regional science models''.} \citep{anselin1988spatial, lesage2009introduction}.  Core methods developed in spatial econometrics account for \textit{spatial dependence}---where the outcome in one location (${y}_i$) influences the outcome in a nearby location (${y}_{i+1}$), and visa versa---as well as spatial spillovers---where the treatment at one location ($\mb{x}_i$) influences the outcome in a nearby location (${y}_{i+1}$)---and numerous tools to measure and model various patterns of spatial autocorrelation of disturbances explicitly \citep{LeSage2010}. In contrast, SFD is simply a research design that exploits the spatial structure of data instrumentally in order to identify the average causal effect of a unit's observable treatments ($\mb{x}_i$) on that same unit's outcome (${y}_i$).  Unlike spatial econometric models, Eq. (\ref{Eq:XsectionSetup}) does not contain any explicit relationships that depend on space, since neither neighbors' treatments nor neighbors' outcomes are on the right-hand side. Space is only used to identify $\beta$ by informing how this equation is transformed via Eq. (\ref{Eq:SFDTheory}). Notably, however, in principle one could write down a model from spatial econometrics and then apply SFD to that model as an identification strategy in the appropriate context.

\paragraph{Models of spatial spillovers}
In many contexts, a treatment at location $i$ may alter an outcome at some nearby location $j$, where the magnitude of the effect decays with distance. For example, construction of a police station may affect crime locally and in nearby locations. Such spatial spillovers are generally accounted for using ``spatial lag models" that recode the \textit{neighbor's treatment} as a new type of treatment variable which is then introduced as an additional independent variable in the original regression. For example, in the model $y_i={x}_i\beta + {x}_{i-1}\gamma+\epsilon_i$, ${\gamma}$ is the spillover from location $i-1$ to $i$. In a more general model, lags in multiple spatial directions and over different distances may be accounted for. Descriptions of spatial lag models and SFD may both draw on language or intuition from time-series\footnote{Descriptions of spatial lag models sometimes invoke similarities to distributed lag time-series models.}, but are employed for unrelated reasons. Spatial lag models may account for a particular data-generating process in which spillovers are present, while SFD is an identification strategy that may be employed regardless of whether spillover are present or not. 

One might imagine that the presence of spatial spillovers necessarily invalidates the the SFD research design because contamination of nearby control groups (i.e. violation of the Stable Unit Treatment Value Assumption) is an obstacle to identification in some spatial settings, such as spatial RD designs, but this is not a problem for SFD. Unlike spatial RD designs, there may be substantial variation in \textit{neighbor's treatments} within a sample so that spatial spillover effects can themselves be identified separately from \textit{own treatment} effects. To see this, note that $i$'s neighbor's treatment (e.g. $x_{i-1}$), which generates a spillover effect to $i$, differs from $i$'s neighbor's {neighbor's treatment} (e.g. $x_{i-2}$), which generates a spillover effect onto $i$'s neighbor. These small differences in adjacent values for the \textit{neighbor's treatment} variable are then exploited in the SFD design to identify the spillover effect.  This intuition generalizes for multiple directions and any distance. Thus, the SFD research design can be used in combination with a spatial lag model without complication if spillovers are thought to be present, all that is required is that the standard spatial lag specification (in levels) is differenced across space.\footnote{
The general spatial lag model is
$$y_i={x}_i\beta + {L}_i(\mb{X})\gamma+\epsilon_i$$
where $\mb{X}$ is the matrix of all treatments for all locations, ${L}_i(.)$ is a spatial lag operator relative to the position of $i$ and $\gamma$ is a vector of estimated spillover effects. Note that ${L}_i(.)$ could encode spatial lags in multiple directions and distances and spillovers need not be isotropic. The SFD model in this general case is then
$$\Delta y_i=\Delta {x}_i\beta + \Delta {L}_i(\mb{X})\gamma+\Delta \epsilon_i.$$
}
 For example, in the simple model of spillovers above, the SFD estimating equation is simply $\Delta y_i=\Delta {x}_i\hat{\beta}_{SFD} + \Delta {x}_{i-1}\hat{\gamma}_{SFD}+ \hat{\Delta\epsilon}_i$ which will recover unbiased estimates of ${\beta}$ and ${\gamma}$. In general, the precise form of spillovers need not be known \textit{ex ante} to achieve unbiasedness, since one can always include ``more than enough'' spatial lags and irrelevant lags will be uncorrelated with the outcome. With sufficient spatial lags included, the orthogonality condition in Eq. (\ref{Eq:IndepCondition}) will be satisfied.

Importantly, however, a mistaken omission of spatial lags in the SFD research design when spatial spillovers are present may generate bias. For example, suppose the data generating process is $y_i={x}_i\beta + {x}_{i-1}\gamma+\epsilon_i$ where $\gamma \neq 0$. If the SFD model is not the model described above but instead omits the spatial lag term, estimated as $\Delta y_i=\Delta{x}_i\hat{\beta}_{SFD}^\dagger +\hat{\Delta\epsilon}_i$, then the estimate $\hat{\beta}_{SFD}^\dagger$ will be biased. This occurs because ${x}_{i-1}$ behaves as a source of unobserved heterogeneity (${x}_{i-1}=\mb{c}_i$ in Eq. \ref{Eq:XsectionSetup}) and the orthogonality condition Eq. (\ref{Eq:IndepCondition}) fails for this particular form of heterogeneity. Unbiasedness of $\hat{\beta}_{SFD}^\dagger$ would require the orthogonality condition $\mathrm{E}[\Delta {x}_{i}\Delta {x}_{i-1}]=2\mathrm{E}[{x}_{i}{x}_{i-1}]-\mathrm{E}[{x}_{i}{x}_{i-2}]-\mathrm{E}[{x}_{i-1}^2]=0$, which will almost never be true; even if there is no spatial autocorrelation in ${x}$, the last term is a variance that will not equal zero. Similar issues arise with spatial spillovers that have other spatial structures. Conceptually, this issue arises because first differencing embeds information regarding $\mb{x}_{i-1}$ into the independent variable $\Delta\mb{x}_i$, with estimated coefficient $\hat\beta_{SFD}$, while at the same time information from $\mb{x}_{i-1}$ persists in the first differenced error term (because of the spillover) if $\Delta\mb{x}_{i-1}$ is not also made a regressor. This contrasts with the standard levels model (Eq. \ref{Eq:XsectionOLS}) because $\mb{x}_{i-1}$ is not incorporated into the regressor with coefficient $\hat\beta_L$, so omission of $\mb{x}_{i-1}$ from the regression does not necessarily induce this mechanical bias.  Notably, though, if there is any spatial auto-correlation in $\mb{x}$---which is almost always true in spatially dense data---and spatial spillovers are present, then estimating the levels regression without spatial lags will also be biased but for a different reason.\footnote{For example, if the data generating process is $ y_i={x}_i{\beta} + {x}_{i-1}\gamma+\epsilon_i$ as above, then estimating a levels models that omits the spatial lag $ y_i={x}_i\hat{\beta}_{L}^\dagger +\hat{\epsilon}_i$ requires the orthogonality condition $\mathrm{E}[{x}_{i}{x}_{i-1}]=0$, which fails if $x$ is spatially auto-correlated.} Thus, inclusion of spatial lags is important for preventing omitted variables bias when spatial spillovers are thought to be important, regardless of whether a cross-sectional model is estimated in levels or SFD.

\paragraph{Spatial autocorrelation robust standard errors}
Recently in applied work, there has been increasing attention to the role of spatial autocorrelation in disturbances when estimating uncertainty in cross-sectional models, particularly in the approach developed by \cite{conley1999gmm}. In Conley's procedure, the structure of spatial autocorrelation is accounted for by computing $\hat\epsilon_i\hat\epsilon_j$ for nearby observations when estimating a covariance matrix, similar to the analogous time-series procedure developed by \cite{newey87}. However, in the SFD research design, all of these cross-covariances in levels are immaterial because common information between units has been differenced out, as discussed above.  Nonetheless, it is possible that the off-diagonal terms in $E[\hat{\Delta \epsilon}\hat{\Delta \epsilon}']$ could be nonzero if there are higher-order aspects of the data-generating process that generate spatial correlations in the gradients of disturbances. In such a scenario, it is appropriate to apply the procedure in \cite{conley1999gmm} to the spatially first-differenced data, which in one-dimensional space is identical to the procedure in \cite{newey87}.  To avoid possible over-rejection of the null, we recommend such an approach when one is unsure about the spatial autocorrelation of $\hat{\Delta \epsilon}$ and we use this approach in empirical examples below. 

\paragraph{Semiparametric regression models}
The SFD design can be understood as a specific case of partially linear semiparametric regression where the vector of unobservable variables $\mb{c}$ is unknown and the dependance of $y$ on $\mb{c}$ is governed by an unknown function $g(\mb{c})$. In its usual formulation, the semiparametric model that would replace Eq. (\ref{Eq:XsectionSetup}) is written $y = \mb{x}\beta + g(\mb{c})+\epsilon$ \citep[e.g.][]{robinson1988root, carroll1997generalized}. We point out that if unobserved covariates $\mb{c}$ are functions of space (with observation $i$ at position $\ell_i$, for consistency with the sections above), then a general solution is to rewrite $g(\mb{c})=g(\mb{c}(\ell_i))=g^*(\ell_i)$ and account for unobservables by estimating a partially linear model that is nonparametric over positions: 
\beq
y_i = \mb{x}_i\beta + g^*(\ell_i)+\epsilon_i.
\label{Eq:PartiallyLinear}
\eeq
Thus, the unobservable cross-sectional heterogeneity described by $\mb{c}_i\alpha$ in Eq. (\ref{Eq:XsectionSetup}) is captured by the nonparametric component $g^*(\ell)$ in the semi-parametric model. SFD leverages the idea that physical space can be used as a metric in which to organize and index observations in order to remove any confounding influence of $g^*(\ell)$.  The SFD solution to estimating Eq. (\ref{Eq:PartiallyLinear}) is to first-difference across space, so that $g^*(\ell)$ is differenced out.  The idea of estimating the linear component of partially linear models through differencing was first proposed by \cite{yatchew1997elementary} based on similar intuition, although, to our knowledge, previous literature has not proposed indexing observations based on physical location for this purpose.

An alternative approach to estimating Eq. (\ref{Eq:PartiallyLinear}) would be the procedure proposed by \cite{robinson1988root}. Implementing Robinson's approach would involve estimating smooth non-parametric ``trends'' in $y$ and $\mb{x}$ across positions $\ell_i$ using kernel estimators, and then regressing the resulting residuals of $y$ on the residuals of $\mb{x}$ in order to recover $\beta$. For example, the ``spatial fixed effects" approach proposed by \cite{conley2010} is a creative implementation of Robinson's approach applied in space, where the kernel used to estimate $g^*(.)$ has uniform mass for all observations within a cutoff radius.\footnote{We demonstrate this implementation in Appendix \ref{AppendixRobinson} and compare the results to SFD for our one-dimensional empirical example.}  Under such an approach, to achieve identification of $\beta$ one would need to assume that all units $j$ near enough to $i$ to inform these kernel estimates at $\ell_i$ are comparable to $i$. This assumption may serve well in some contexts, but may be difficult to defend if the elements of $\frac{\partial \mb{c}}{\partial \ell}$ (and thus possibly $\frac{\partial g^*}{\partial \ell}$) are hypothesized to be highly variable across space, to exhibit unknown discontinuities, or to be anisotropic in the neighborhood of $\ell_i$. SFD circumvents this assumption, which may be useful if, for example, crucial dummy variables that would otherwise capture level-shifts in $g^*(\ell)$ were omitted from the model (in the SFD approach, effects of omitted dummy variables will nonetheless be absorbed even if they are not explicitly accounted for). Differencing can be thought of as restricting the bandwidth of the kernel estimator in the first stage of Robinson's procedure to its very smallest possible value, such that it contains only a single neighboring observation. In this sense, one could think of SFD ``as if'' it applies Robinson's procedure by estimating $g^*(\ell)$ using a completely nonparametric ``spatial trend'' that includes a single dummy variable for every single pair of neighboring observations. However, critically, implementing such a procedure using dummy variables is not identified for $N$ adjacent observations, since it would require estimating at least $N$ parameters ($N-1$ dummy variables and $\hat{\beta}$). Thus, SFD is the only feasible approach that also allows for this level of flexibility in the possible structure of $g^*(\ell)$. Although, notably, the cost of this flexibility is a loss of efficiency.  As shown by \cite{yatchew1997elementary}, Robinson's approach converges faster. 

%-------------------------------------PRACTICAL CONSIDERATIONS

\section*{Practical considerations for the plausibility of identification}

The design of the SFD estimator emerges naturally from recognizing that adjacent neighbors in a sample may be comparable, although exact comparability across every single pair of neighbors is not actually essential for SFD to provide identification.  If the Local Conditional Independence assumption (Eq. \ref{Eq:LocalUnitHomogeniety}) is true, then the identifying assumption that $\Delta \mb{x}$ and $\Delta \mb{c}$ are orthogonal (Eq. \ref{Eq:IndepCondition}) will hold. However, Eq. (\ref{Eq:IndepCondition}) may remain valid under weaker conditions in actual data. Here we discuss some practical issues to consider when determining whether SFD may provide causal identification in different contexts.

 \paragraph{The spatial density of observation}
The assumption that adjacent observations in a data set are comparable relies on the notion that observations that are adjacent in a data set are actually ``nearby'' one another in the real world. If adjacent observations in a data set are extremely far from one another in actual space, then it may not be reasonable to assume they are comparable. For example, in a sample of countries, China and Russia are adjacent, but from the standpoint of many economic questions, they may not be directly comparable. One reason they do not seem comparable, intuitively, is that there are many portions of Russia that are extremely distant from many portions of China, despite the existence of a common boundary between the two units. Thus, when assessing whether SFD is an appropriate approach for a given sample, it is important to consider whether adjacent observational units are nearby one another in their entirety, especially with respect to the range of spatial coverage across the sample.  For this reason, it is likely that local conditional independance will be most nearly true, and SFD well identified, if the spatial extent of individual observational units is limited and their spatial density is high. We cannot provide precise guidance on how dense is ``dense enough'' in practice, rather, it is up to the judgement of the econometrician to determine whether the spatial density of observations is sufficiently high that Eq. (\ref{Eq:LocalUnitHomogeniety}) (or at least Eq. \ref{Eq:IndepCondition}) is likely to be satisfied.
  
 \paragraph{Considering potential common causes of both regressors and omitted variables}

\begin{figure}%[b!]
\begin{center}
\includegraphics[width=\textwidth]{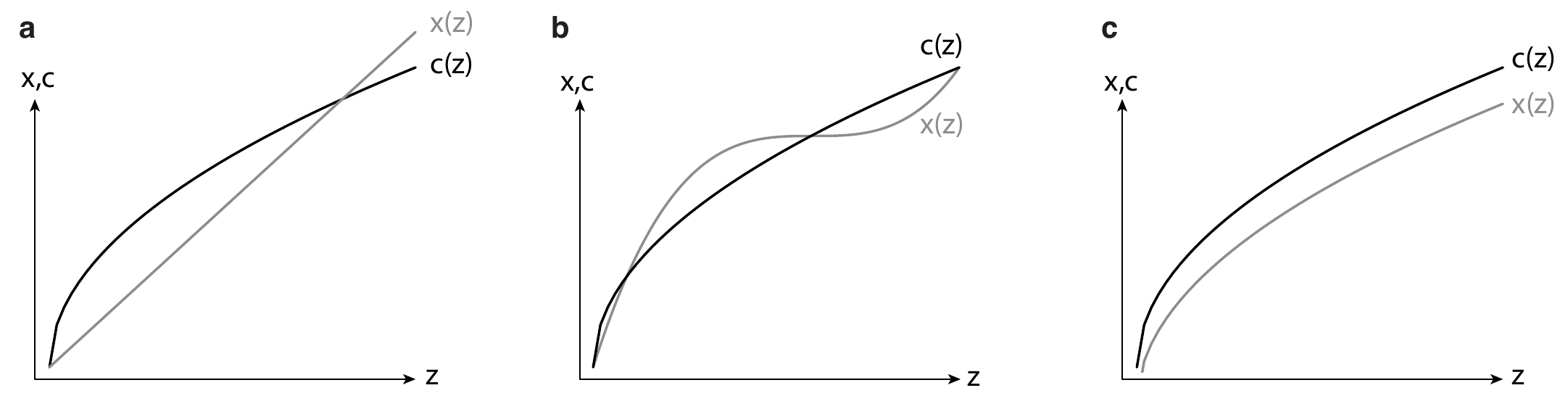}
\caption{{\bf Scenarios under which a common cause would and would not generate bias.} Cases where both an observed variable $x$ and unobserved covariate $c$ are affected by a common unobserved cause $z$. Panels (a) and (b) satisfy Eq. (\ref{Eq:CommonCause}) since second derivatives in $z$ are unrelated, leading first derivatives to be uncorrelated such that $\hat{\beta}_{SFD}$ is unlikely to be unbiased. Panel (c) generates correlated first derivatives because the curvature of $x(z)$ mirrors that of $c(z)$, causing bias in $\hat{\beta}_{SFD}$.
}
\label{Fig:CommonCause}
\end{center}
\end{figure}

A natural question to consider is whether the SFD estimator will be identified in cases where some unobserved variable $z$ influences both the regressors $\mb{x}$ and the unobserved variable $\mb{c}$:
$$\mb{x}_i = \mb{x}(z_i),\;\;\; \mb{c}_i = \mb{c}(z_i).$$
In this case, as the sample is traversed in physical space, $z$ will evolve as a function of location, driving changes in both $\mb{x}$ and $\mb{c}$ along the way. For example, higher elevations ($z$) across counties might cause temperatures (${x}$) to fall and the air to be thinner (${c}$), both of which might in turn affect crop yields ($y$).  In a levels model, the existence of such an external factor $z$ might generate correlation in $\mb{x}$ and $\mb{c}$, thereby inducing bias in $\hat{\beta}_{L}$. However, $\hat{\beta}_{SFD}$ is substantially more robust to this scenario, in the sense that such a bias is much less likely even if an unknown common cause exists, because a violation of the identifying assumption occurs only if a fairly restrictive condition on the functions $\mb{x}(z)$ and $\mb{c}(z)$ is met.  Specifically, a common cause $z$ generates bias if the \textit{curvature} of the functions $\mb{x}(z)$ and $\mb{c}(z)$ mirror one another throughout the support of $z_i$ in a sample. For example, if $\mb{c}(z)$ is concave in $z$ and $\mb{x}(z)$ increases linearly in $z$ (Figure \ref{Fig:CommonCause}a) or exhibits higher order variations as a function of $z$ (Figure \ref{Fig:CommonCause}b), then it is likely SFD will be unbiased---even though $\mb{x}$ and $\mb{c}$ are correlated in levels across $z$. 

To see why this is true in general, note that for sufficiently small changes in $z$ between neighbors we obtain the first-order Taylor approximations
$$
\Delta \mb{x}_i = \frac{\partial \mb{x}(z_{i-1})}{\partial z} \Delta z_i, \;\;\; \Delta \mb{c}_i = \frac{\partial \mb{c}(z_{i-1})}{\partial z} \Delta z_i.
$$
We use these approximations to rewrite the local conditional independence condition in Eq. (\ref{Eq:IndepCondition}) as
\beq
E[(\Delta\mb{x}_i-\overline{\Delta\mb{x}})'\Delta\mb{c}_i]=E\left[\left(\frac{\partial \mb{x}(z_{i-1})}{\partial z}-\overline{\frac{\partial \mb{x}(z_{i-1})}{\partial z}} \right) '  \frac{\partial \mb{c}(z_{i-1})}{\partial z} \Delta z_i^2\right]=0_{K,M}
\label{Eq:CommonCause}
\eeq
where, for clarity, we explicitly write out the subtraction of the sample average of $\Delta\mb{x}$, denoted $\overline{\Delta\mb{x}}$,  which is partialled out by the vector of ones in the SFD regression.  If $\Delta z_i$ is held fixed\footnote{Note that for vanishingly small changes in physical position $\Delta \ell \rightarrow 0$ , $\Delta z$ is locally constant ($\Delta z = \frac{\partial z}{\partial \ell}\Delta \ell$ by Taylor's theorem).}, this expression indicates that SFD will fail to be identified only if the demeaned derivative of a regressor with respect to the common cause ($\frac{\partial x}{\partial z}-\overline{\frac{\partial x(z_{i-1})}{\partial z}}$) is correlated with the derivative of the omitted variable with respect to the common cause ($\frac{\partial c}{\partial z}$) across the values of $z$ in the sample. In order for such a violation to occur, the \textit{second derivatives} of $\mb{x}(z)$ and $\mb{c}(z)$ must correspond across the support of $z$, thereby generating correlation between the first derivatives in Eq. (\ref{Eq:CommonCause}). Such a situation is shown in Panel c of Figure \ref{Fig:CommonCause}, where Eq. (\ref{Eq:CommonCause}) is likely violated. In this case, as space is traversed and $z_i$ varied, changes in $\Delta\mb{x}-\overline{\Delta\mb{x}}$ and changes in $\Delta\mb{c}$ will occur at exactly the same positions, leading to correlation in spatial first differences of these variables. Nonetheless, in most contexts, we believe scenarios similar to Panels a and b of Figure \ref{Fig:CommonCause} are generally more likely for most common causes one might postulate.

In practice, to satisfy Eq. (\ref{Eq:IndepCondition}), it is generally sufficient for the curvature of $\mb{x}(z)$ and $\mb{c}(z)$  to differ over the range of $z$ plausibly contained within a sample, not everywhere in $z$.  Of course, it is also possible for Eq. (\ref{Eq:CommonCause}) to be violated if the square of the first difference in $z$ is correlated with the product of the two derivatives, although we think such conditions are relatively exotic for most applied settings. 

\paragraph{Rotation of coordinates}

Because $\mb{c}$ is unobserved, it is impossible to directly test whether Eq. (\ref{Eq:IndepCondition}) holds. Nonetheless, we are able to offer a practical indirect check that is implementable using the data. We suggest ``cross-checking'' the identifying assumption of SFD across multiple implementations of the estimator. In Figure \ref{Fig:Implementation}, we pointed out that in two-dimensional environments, SFD could be implemented in the East-West direction or the North-South direction, providing two separate estimates of $\hat{\beta}_{SFD}$ that can be compared to one another for consistency. Results that match would suggest that either the East-West and North-South versions of Eq. (\ref{Eq:IndepCondition}) both are true, or they both fail and somehow generate bias of similar structure despite exploiting different sources of variation in $\Delta\mb{x}$.  We point out that in many environments, this intuition can be further generalized by noting that if Eq. (\ref{Eq:IndepCondition}) holds in general, an econometrician should be able to estimate SFD by taking differences along an axis that has been rotated in space by an arbitrary angle $\theta$ and the resulting estimate $\hat{\beta}_{SFD}(\theta)$ should be relatively invariant across $\theta$. In our analysis of US maize yields below, we demonstrate this test for 180 estimates of key parameters as the coordinate system we use is rotated through $\theta=-89^\circ$ to  $\theta=90^\circ$ by $1^\circ$ increments.

\paragraph{Spatial Double Differences}

Another indirect check that is implementable using the data and which requires different identifying assumptions involves taking higher order differences. Differencing Eq. (\ref{Eq:SFDTheory})
\begin{align*}
 \Delta y_i - \Delta y_{i-1} &= (\Delta \mb{x}_i \beta + \Delta \mb{c}_i \alpha+ \Delta \epsilon_i) - (\Delta \mb{x}_{i-1} \beta + \Delta \mb{c}_{i-1} \alpha+ \Delta \epsilon_{i-1})\notag\\
\Delta^2 y_i&=\Delta^2 \mb{x}_i \beta + \Delta^2\mb{c}_i \alpha + \Delta^2 \epsilon_i
\label{Eq:DFDTheory}
\end{align*}
we obtain the double difference of Eq. (\ref{Eq:XsectionSetup}) where $\Delta^2 \mb{x}_i = \Delta \mb{x}_i - \Delta \mb{x}_{i-1}$. Thus we propose the ``Spatial Double Differences" (SDD) cross-sectional regression model
\beq
\Delta^2 y_i=\Delta^2 \mb{x}_i \hat{\beta}_{SDD}  + \widehat{\Delta^2 \epsilon_i}
\label{Eq:SDDModel}
\eeq
which can be estimated via OLS and will be unbiased so long as
\beq
E[\Delta^2\mb{x}'\Delta^2\mb{c}]=0_{K,M}
\label{Eq:DFDIndependance}
\eeq
which is a distinct restriction that is neither implied by nor a result of Eq. (\ref{Eq:IndepCondition}). We suggest that in cases where one is uncertain that Eq. (\ref{Eq:IndepCondition}) is true, an econometrician might estimate Eq. (\ref{Eq:SDDModel}) and compare whether $\hat{\beta}_{SDD}=\hat{\beta}_{SFD}$. If these estimates are very close then, similar to the comparison with a rotated coordinate system, in order for $\hat{\beta}_{SFD}$ to be biased by a failure of Eq. (\ref{Eq:IndepCondition}), one must postulate a structure for $\mb{c}$ such that $\hat{\beta}_{SDD}$ is identically biased by the failure of Eq. (\ref{Eq:DFDIndependance}). If such a situation is deemed unreasonable, then it is likely that both Eq. (\ref{Eq:IndepCondition}) and Eq. (\ref{Eq:DFDIndependance}) are true. Given sufficient data, simultaneous failure of these conditions but identical SFD and SDD estimates is difficult to achieve in practice, leading us to argue that this is a strong test.  However, in many data environments, it maybe be challenging to estimate $\hat{\beta}_{SDD}$ because the variation in double-differenced variables is likely to be noisy, which can lead to imprecise estimates of $\hat{\beta}_{SDD}$ and attenuation bias. Nonetheless, we demonstrate implementation of this test below in our analysis of US maize yields. 
 
%-------------------------------------SIMULATION + RETURNS TO SCHOOLING

\section*{Comparisons of omitted variables biases in one-dimensional examples}

The core benefit of the SFD research design is elimination of omitted variables bias induced by spatial correlations between regressors and omitted variables in cross section. The magnitude of this benefit is described by Eq. (\ref{Eq:OVBComparison}), computed by comparing $\hat{\beta}_L$ and $\hat{\beta}_{SFD}$. In general, this difference will be dominated by the unobservable $\mb{W}_1$ term. In this section, we provide simple examples in one-dimensional space that demonstrate the magnitude of this benefit by comparing cross-sectional regression results using the standard levels model and SFD. We first consider an idealized simulation using synthetic data where the structure of the omitted variable bias is known exactly. Then we consider estimates for the returns to schooling in census blocks along 10th Avenue in New York and I-90 in Chicago, where omitted variables are not known but plausible values of $\beta$ have been well-documented and replicated in prior analyses, providing a benchmark for comparison.  

\subsection*{An idealized simulation}

\begin{figure}
\vspace*{-1cm}
\begin{center}
\includegraphics[scale=0.7]{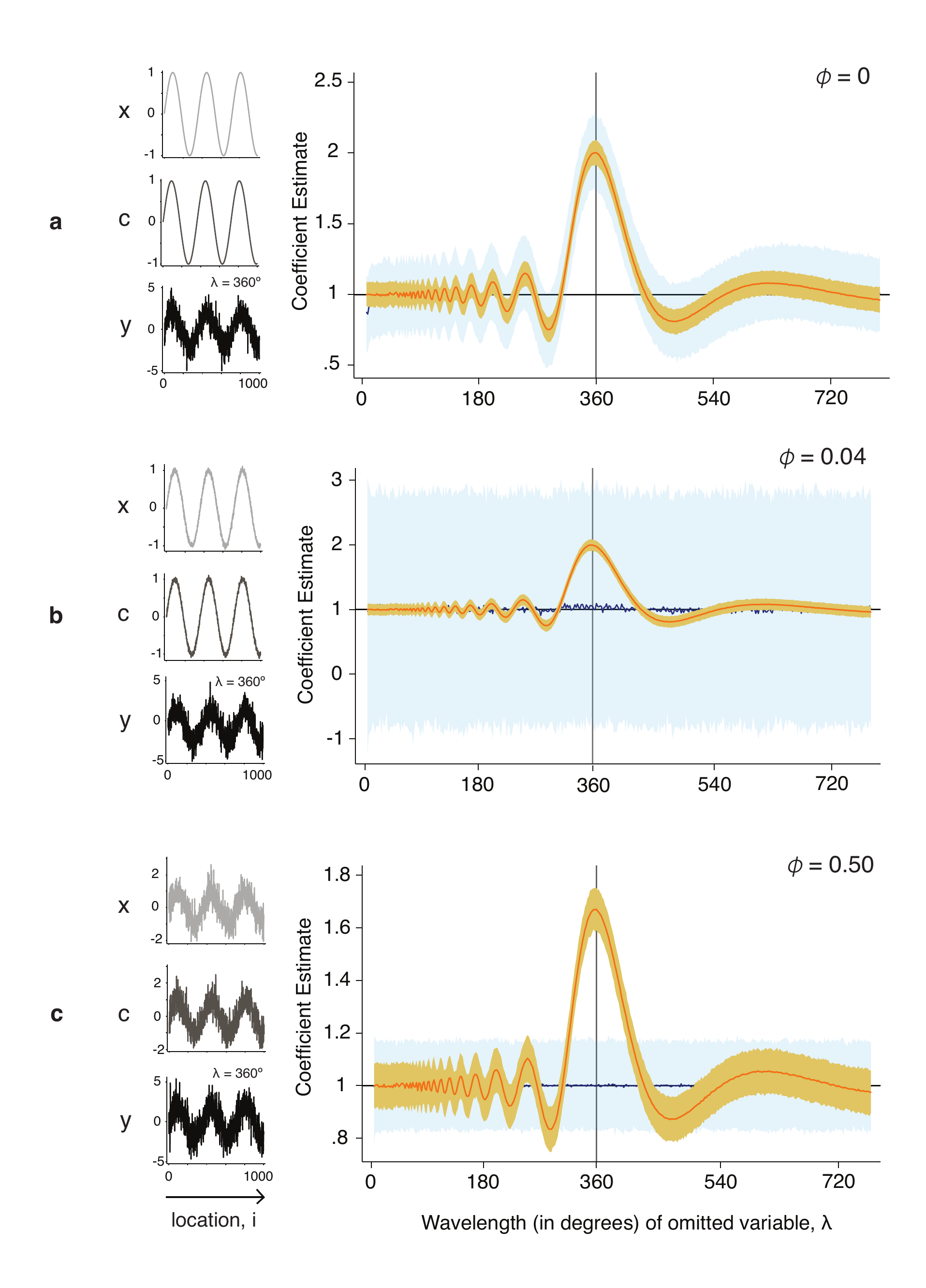}
\caption{{\bf Performance of OLS in levels and first differences in the presence of an omitted variable.} Each panel shows the same simulation, but with three different levels of noise ($\phi$) added to the independent variables. The three plots on the left show an example of the explanatory variable x (observed), the omitted variable c (not observed) and the corresponding outcome variable y (observed) when $\lambda = 360$. The plots on the right show the coefficient estimates (true value = 1) as a function of $\lambda$, the wavelength of the sinusoidal component of the omitted variable. Levels estimates are shown in orange and SFD estimates are shown in blue. 
The darker lines show the average coefficient estimates, and the lightly shaded areas show the inner 95\% of estimates. The sample size is 1,000. 1,000 repetitions were run for each parameterization of the problem $\{ \phi , \lambda \}$. (a) $\phi = 0$. (b) $\phi = 0.04$. (c) $\phi = 0.50$.}
\label{Fig:Simulation}
\end{center}
\end{figure}

In this simple simulation, we generate synthetic data to compare the performance of SFD to that of levels in the presence of a single, known omitted variable. This exercise demonstrates some conditions under which the SFD estimator performs well. The data generating process is as follows. Let $i = 1, \dots, 1000$ index evenly spaced observations along a line (note that here, $i = \ell_i$). We are interested in the outcome variable $y$ that is determined by $x$, which is observed, and $c$, which is not observed: 
$$
y_i = x_i\beta +  c_i\gamma + \epsilon_i
\label{Sim_Model}
$$
where $\beta = \gamma = 1$ and $\epsilon_i \sim N(0,1)$. In order to allow us to smoothly vary the degree of spatial correlation between $x$ and $c$, we exploit sinusoidal functions (in degrees, not radians) and vary their wavelength. Specifically, we generate 
$$
x_i = \mathrm{sin}(i) + \delta_i  \phi;
\label{Sim_X}
$$
$$
c_i = \mathrm{sin}\Big( \frac{360 i }{\lambda}\Big) + \eta_i \phi ; 
\label{Sim_C}
$$
where $\delta_i$ and $\eta_i$ are disturbance terms that are both independently distributed $N(0,1)$. Throughout the simulation, the expected value of $x$ completes one cycle every $360$ observations, so its wavelength is $360$. The wavelength of $c$ is controlled by $\lambda$, such that $x$ and $c$ are most highly correlated when $\lambda = 360$. The noise terms, $\delta_i$ and $\eta_i$, are amplified by the parameter $\phi$, which we will also vary. We run 1,000 repetitions for each parameterization of the problem, defined by the values of $\lambda$ and $\phi$. The three subplots to the left in Figure \ref{Fig:Simulation} show the explanatory variable $x$, the omitted variable $c$, and the outcome variable $y$ for the parameterization $\lambda = 360$ when $\phi = 0$ (panel a), $\phi = 0.04$ (panel b), and $\phi = 0.5$ (panel c).

Since $c$ is unobserved, we estimate $\beta$ (true value = 1) by regressing $y$ on $x$ and intentionally omit $c$ from the model. For each simulation we estimate the levels model 
$$
y_i = \hat{\alpha}_1 +  x_i\hat{\beta}_L +  \hat{\epsilon}_i
\label{Sim_Levels}
$$
and the SFD model
$$
\Delta y_i = \hat{\alpha}_2 +  \Delta x_i\hat{\beta}_{SFD} + \hat{ \Delta \epsilon}_i
\label{Sim_SFD}
$$
and record $\hat{\beta}_L$ and $\hat{\beta}_{SFD}$.

The results for three values of $\phi = \{0, 0.04, 0.5\}$ and integer values of $\lambda = \{1, \dots, 800\}$ are displayed in the right panels of Figure \ref{Fig:Simulation}. Each subplot displays the coefficient estimates as a function of $\lambda$, with the levels estimates shown in orange and the SFD estimates shown in blue. The lightly shaded areas show the inner 95\% range of estimates over the 1,000 simulations, while the darker lines show the average across all 1,000 estimates. In the cases with no noise ($\phi = 0$; panel a), the mean $\hat{\beta}_{SFD}$ is very near to the mean $\hat{\beta}_{L}$ because $x$ and $c$ are perfectly correlated. Both estimators perform well for small values of $\lambda$, since at these values $x$ and $c$ have a low degree of spatial correlation. However, the two estimators perform poorly for $\lambda = 360$ because the observed variable and the omitted variable are perfectly correlated across space. In this situation, all the influence of omitted variable $c$ is picked up by $\hat{\beta}$ in both levels and SFD, creating large biases.  

However, with even a minute amount of noise, $\Delta x$ and $\Delta c$ become uncorrelated in expectation and the bias becomes considerably smaller for the SFD estimate than for the levels estimate. This result is demonstrated in panel b of Figure \ref{Fig:Simulation}. When $\phi = 0.04$ (panel b), the bias of $\hat{\beta}_{SFD}$ is essentially zero, whereas the absolute bias for $\hat{\beta}_L$ is 0.08, on average across $\lambda$. An important trade-off in this case is that the efficiency of $\hat{\beta}_{SFD}$ relative to $\hat{\beta}_{L}$ is low ($\frac{Var(\hat{\beta}_{L})}{Var(\hat{\beta}_{SFD})}=0.08$). However, as the amplitude of noise increases, SFD remains less biased than levels, but its variance declines substantially relative to the levels estimator. This relative decline occurs because the presence of noise in $x$ increases the variation that can be exploited in first differences, reducing $(\Delta x' \Delta x)^{-1}$. When the quantity of noise is modest ($\phi = 0.5$; panel c), the bias of $\hat{\beta}_{SFD}$ is less than 0.2\% that of $\hat{\beta}_{L}$, on average across $\lambda$, and the relative efficiency of $\hat{\beta}_{SFD}$ to $\hat{\beta}_{L}$ is much better at 0.5. 

The difference between the SFD estimate and the levels estimate is most striking when the variable of interest and the omitted variable are highly spatially correlated, for the cases with non-zero $\phi$. While the levels estimator always performs badly when $\lambda$ is in the neighborhood of 360, such that $x$ and $c$ are perfectly in phase (as shown in the small subplots), the SFD estimator is remarkably unbiased, performing no worse than for other $\lambda$. For example, when $\lambda = 360$ and $\phi = 0.5$, the average value of $\hat{\beta}_{SFD}$ is $1.00$ (95\% interval: $0.84 \leq \hat{\beta}_{SFD} \leq 1.16$), while the average value of $\hat{\beta}_{L}$ is $1.67$ (95\% interval: $1.59 \leq \hat{\beta}_{L} \leq 1.75$).

This simple exercise illustrates the conditions under which the SFD estimator performs well. When the regressor of interest and the omitted variable are highly spatially correlated, the SFD estimator is remarkably less biased than the levels estimate. When these two variables are not as strongly correlated across space, the bias in SFD is no worse than levels. There is a tradeoff between bias and efficiency though, so the SFD estimate tends to have a larger variance. Nevertheless, so long as there is sufficient orthogonal variation in the independent variables, the efficiency of the SFD estimator may be comparable to that of the levels estimator. Next, we compare the results from the two estimators in a one-dimensional empirical example where plausible parameter values have been previously established.

 \subsection*{Returns to schooling along 10th Avenue and Interstate-90}
 
In real world contexts, we do not observe omitted variables, making it impossible to know for certain how close $\hat{\beta}_{SFD}$ is to $\beta$. But what we can do, at least for an illustrative example, is estimate $\hat{\beta}_{SFD}$ for a relationship that is sufficiently well studied that we have some sense of the true value for $\beta$. In the following example, we demonstrate how SFD is implemented in one-dimensional space by conducting a simple analysis similar to the returns to schooling example discussed above, examining census blocks along the longest roads in Manhattan and Chicago. We then compare the $\hat{\beta}_{SFD}$ and $\hat{\beta}_{L}$ estimates with these samples to previous estimates of the returns to education. 

\begin{figure}
\vspace*{-1cm}
\begin{center}
\includegraphics[width = \textwidth]{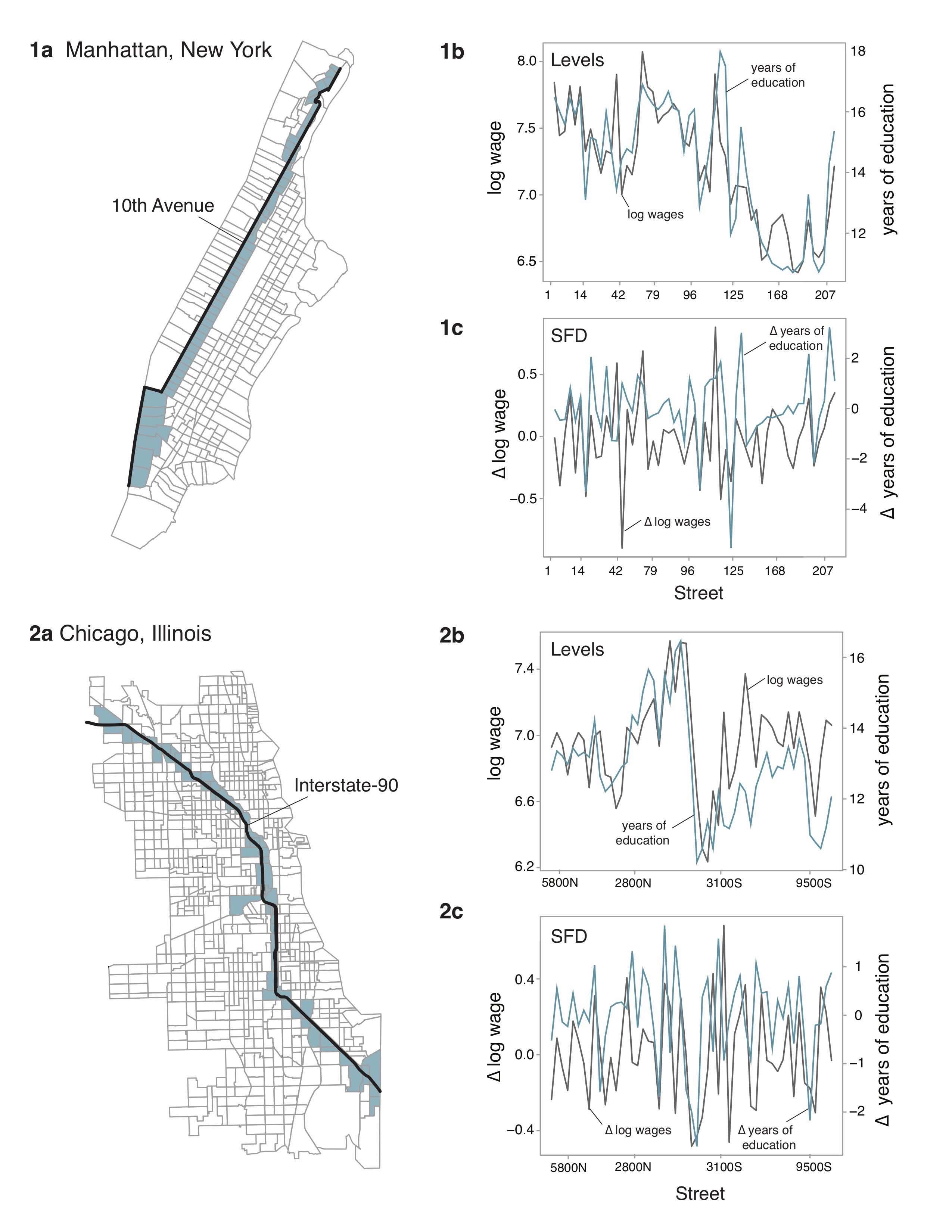}
\caption{{\bf Implementation of SFD along 10th Ave and I-90 to estimate returns to education.} (1a) and (2a) show the sequence of census blocks included each sample (blue), which lie along the longest roads (black) in Manhattan, New York City and Chicago. (1b) and (2b) display log weekly wages (grey) and years of schooling (blue) in the New York and Chicago samples, respectively. (1c) and (2c) are the same data after taking SFD.}
\label{Fig:ReturnsToSchooling}
\end{center}
\end{figure}

We obtain data on average wages and average years of education from the 2010 American Community Survey 5-year estimates for New York City and Chicago \citep{ACSdataset}. In New York, we produce a sample of 53 adjacent observations by following 10th Avenue from the lower to the upper tip of Manhattan and recording the census tracts along this path, as depicted in Figure \ref{Fig:ReturnsToSchooling} (panel 1a). We follow the same procedure in Chicago, tracing Interstate-90 from the northwestern to the southeastern corner of the city and obtaining 54 sequential observations (panel 2a). We arrange the observations in order, such that we have a one-dimensional sequence of adjacent census blocks. Panels 1b and 2b of Figure \ref{Fig:ReturnsToSchooling} show the levels of log weekly wages (grey) and years of schooling (blue) that would be observed driving north along 10th Avenue in Manhattan and southeast along I-90 in Chicago, respectively. For comparison, Panels 1c and 2c show the spatial first differences in log weekly wages (grey) and years of schooling (blue) along the same paths. 

Indexing census tracts by $i$ according to their position along these two roads, we estimate the effect of years of education on wages via OLS, intentionally omitting all covariates. The levels model is
\beq
log(wage_i) = \hat{\alpha}_1 + years\_of\_education_i \hat{\beta}_{L} + \hat{\epsilon_i}
\label{RTS_levels}
\eeq
and the SFD model is
\beq
\Delta log(wage_i) = \hat{\alpha}_2 + \Delta years\_of\_education_i \hat{\beta}_{SFD} + \hat{\Delta \epsilon_i}
\label{RTS_diffs}
\eeq
where $\hat{\alpha}_1$ is the intercept in the levels model and $\hat{\alpha}_2$ is the intercept in the SFD model.\footnote{We estimate returns to schooling on this dataset using the semi-parametric model proposed by Robinson (1998) in Appendix \ref{AppendixRobinson}.}

 \begin{table}
\thispagestyle{empty}
\fontsize{9}{9}\selectfont
\begin{center}
\newcolumntype{Y}{>{\centering\arraybackslash}X}
\begin{tabularx}{\linewidth}{ l Y Y Y Y Y Y}
\hline
\hline
 & & & & & & \\
 & \multicolumn{6}{c}{\textit{Dependent variable: log average wage}} \\
 & & & & & & \\
 \cline{2-7} 
 & & & & & & \\
 & \multicolumn{2}{c}{10th Avenue, New York} & \multicolumn{2}{c}{I-90, Chicago} & \multicolumn{2}{c}{Staiger and Stock (1997)} \\
 & & & & & & \\
   & Levels & SFD & Levels & SFD & OLS & IV \\
   & & & & & & \\
\hline
 & & & & & & \\
 Average years of education  & $0.178$*** & $0.089$** & $0.124$*** & $0.072$* & $0.063$*** & $0.098$***  \\ 
 & $(0.015)$ & $(0.028)$ & $(0.020)$ & $(0.037)$ & $(0.000)$ & $(0.015)$ \\
 & & & & & & \\
Constant & $4.682$*** & $-0.007$ & $5.355$*** &  $0.000$ & -- & -- \\
 & $(0.217)$ & $(0.040)$ & $(0.259)$ & $(0.035)$ & & \\
  & & & & & & \\
 \hline
 & & & & & & \\
 Observations & $53$ & $52$ & $54$ & $53$ & $329,509$ & $329,509$ \\
 R squared & $0.73$ & $0.16$ & $0.43$ & $0.07$ & -- & -- \\
  & & & & & & \\
 \hline
  \hline
  \end{tabularx}
 \caption{{\bf Cross-sectional estimates for returns to education using levels and SFD.} Data for the first four columns are for census tracts in Manhattan, New York along 10th Avenue and Chicago, Illinois along Interstate-90 for the year 2010. We report OLS standard errors, which, in this case, are more conservative than Newey-West standard errors. Asterisks indicate statistical significance at the 0.1\%,***, 1\%**, and 5\%* levels.}
 \label{Tab:ReturnsToSchooling}
 \end{center}
 \end{table}
 
The results are shown in Table \ref{Tab:ReturnsToSchooling}. The semi-elasticities estimated using levels are 0.179 in Manhattan and 0.125 in Chicago, suggesting that each additional year of schooling increases wages by 18\% and 12.5\% in these cities, respectively. These values are larger than almost all previous estimates of the return to education. Card (2001) reports 17 previous estimates of the return to education in the United States, which range from 0.052 to 0.132 \citep{card2001estimating}. The levels estimate in Manhattan is much larger than all of these estimates, and the estimate in Chicago is larger than all but one. In contrast, the coefficients we estimate using SFD, 0.089 in New York and 0.072 in Chicago, are in the center of the distribution of previous estimates. For comparison, the OLS and IV estimates from \cite{staiger1997instrumental}, the largest study reviewed in Card (2001), are also reported in Table \ref{Tab:ReturnsToSchooling}. 

The difference between the levels and SFD estimates reported in Table \ref{Tab:ReturnsToSchooling} is the magnitude of the omitted variables bias that is eliminated from the cross section by differencing out spatial histories, as shown in Eq. (\ref{Eq:OVBComparison}). As displayed clearly in Figure \ref{Fig:ReturnsToSchooling}, panels 1b and 2b, the correlation in spatial histories ($\mb{W}_1$) is large. The positive sign of this difference seems sensible when considering which omitted variables are likely to generate bias in the levels estimate. For example, if households that have high levels of education tend to live in areas with more Whites and Whites tend to earn more than other races, then race will be a spatially correlated omitted variable that will lead to upward bias in the levels estimate.

This analysis demonstrates that we recover well established estimates of the return to education using SFD even when {all} covariates are intentionally omitted from the analysis. In two different cities, with a modest sample size determined only by the location of the longest road, we produced estimates of the return to education that appear precise and match previous estimates. 
    
%-------------------------------------EMPIRICS: CROPS

\section*{Empirical application: effects of geographic factors on crop yields}

We now develop new cross-sectional estimates for the effect of long-term climate and average soil conditions on maize yields in US counties, relationships notoriously confounded by unobserved heterogeneity, using SFD. Further, this rich data set allows us to explore three additional points: (i) implementation of SFD using irregular (non-gridded) two-dimensional data; (ii) systematic evaluation of the research design's vulnerability to omitted variables bias using an ``extreme bounds" analysis \citep{leamer1985sensitivity}; and, (iii) implementation of two novel robustness tests unique to SFD, the rotation of the coordinate system and SDD, which enable us to check the research design's underlying assumptions.

\subsection*{Context}

Understanding the impact of soil and climatic endowments on economic outcomes has become an important area of research in development economics, economic geography, and environmental economics, motivated by strong cross-sectional patterns of development and, more recently, climate change \citep[e.g.][]{bloom1998geography, acemoglu2000colonial, easterly2003tropics,  nordhaus2006geography, carleton2016social}.  However, analyses of economic responses to long-run geographic conditions, such as average temperature, are potentially affected by omitted variables bias since comparing outcomes in areas with different conditions directly (e.g. hot vs. cold locations) has long employed cross-sectional research designs.  In this literature, unobserved heterogeneity has been identified as a key issue in particular, as climate variables are thought to be spatially correlated with other variables (e.g. ruggedness, political institutions) that may not be observed but likely affect economic outcomes \citep{deschenes2007economic, hsiang2016climate}. As with other cross-sectional contexts, there is no systematic way to determine whether a key variable has been omitted. Nonetheless, prior analyses have worked to address the omitted variables issue by saturating the levels model with covariates \citep[e.g.][]{nordhaus2006geography} or assuming confounding factors are orthogonal to climate \citep[e.g.][]{acemoglu2000colonial}. 

An alternative widely-used approach is to estimate a panel regression model that exploits random inter-temporal variation in environmental variables and includes location-specific fixed effects, which partial-out unobserved cross-sectional differences that influence the outcome of interest. In the context of climate effects, this approach has been used to estimate the effect of high-frequency weather variation on seasonal crop yields \citep{deschenes2007economic, schlenker2009nonlinear, lobell2011climate, auffhammer2014empirical}, marginal effects that are then convolved with climate-specific weather distributions to reconstruct estimates for the effect of climate. However, this approach has remained contentious because it is unclear whether marginal effects of weather can be assumed similar to the marginal effects of changing the climate once populations have fully adapted \citep{hsiang2016climate}.  Furthermore, this approach cannot be applied to understand the effect of other environmental factors, such as soil quality, that exhibit little or no inter-temporal variation within a location on practical time scales. To our knowledge, the long-run effects of time-invariant natural factors, such as soil quality, remain essentially unstudied empirically in modern economics because the challenge of unobserved heterogeneity has been insurmountable.\footnote{\cite{hornbeck2012nature} is an important and notable exception.}

Here, we demonstrate that SFD presents an appealing ``third path'', allowing us to estimate the impact of long-run geographic conditions while eliminating the effects of unobserved heterogeneity.  Our estimates can be thought of as the effect of long-run climate and soil quality net of adaptation and all long-run adjustments, in the sense originally articulated by \cite{mendelsohn1994impact}.  To evaluate the research design's performance in the presence of omitted variables, we again compare SFD estimates to standard levels estimates when covariates are intentionally and systematically withheld from the regression model. 

The remainder of this section is organized as follows. First, we introduce our data and cross-sectional specification. Second, we estimate the effect of climate and soil on maize yields using the levels research design. Third, we demonstrate how SFD can be implemented with irregular units in two-dimensional space and estimate these effects via SFD, recovering new estimates for the effect of long-run climate and soil on yields net of unobserved heterogeneity. Fourth, we systematically withhold covariates from both regression models to compare how the two research designs perform in the presence of omitted variables. Lastly, we conduct two internal robustness checks for the SFD approach: a continuous rotation of the coordinate system and SDD. 

 \subsection*{Data}
 
We obtained the data on annual county-level maize yield, temperature, and rainfall for the years 1950-2005 used in \cite{schlenker2009nonlinear} and the vector of five soil characteristics used in \cite{schlenker2006impact}. The soil characteristic include the \textit{minimum permeability} (inches per hectare), \textit{average water capacity} (inches per inch), \textit{soil erodibility factor} (0.02 for the least erodible soils to 0.64 for the most erodible), \textit{percent clay content} (\%), and \textit{percent of high-class top soil} (\%). Following these authors, we limit our analysis to the balanced panel of counties east of the 100th meridian. To demonstrate the benefits of SFD in cross-sectional data, we average the weather and yield data over the 56-year period, so there is only one observation per county. We then match these datasets with the soil data, which is already a cross section, to create a final cross section. Importantly, our long-term averages of weather (temperature and rainfall) can be viewed as measures of local climate defined in earlier work by \cite{mendelsohn1994impact}. To the best of our knowledge, effects of cross-sectional variation in soil conditions have not been a focus of prior study, perhaps in part due to concerns of unobserved heterogeneity, although changes in soil quality over time have been analyzed \citep[e.g.][]{hornbeck2012nature}. 

\subsection*{Specification}
 
We employ a log-linear regression model with maize yield as the dependent variable. There are nine explanatory variables describing seven environmental conditions, since nonlinear effects of temperature and precipitation are each described by two variables.

Using data from \cite{schlenker2009nonlinear}, we represent the non-linear effect of temperature on maize yields using a formulation that reflects a piecewise-linear spline in hourly temperatures during the growing season. This approach measures the amount of time a crop is exposed to various temperatures at high temporal resolution, but collapsed so that it may be matched to yield data that is collected after longer intervals of time over which crop growth occurs. Our specification allows the effect of hourly temperature to have a different marginal effect depending on whether the temperature is above or below 29\degree C. \textit{Degree-days below 29\degree C} is a variable whose coefficient captures the effect on end-of-season yields from a marginal 24 hour period at temperatures between 0 and 29\degree C. The coefficient for the \textit{degree-days above 29\degree C} variable describes the effect of a marginal 24 hour period at temperatures above 29\degree C.  \cite{schlenker2009nonlinear} established that large declines in maize yield occur for temperatures above 29\degree and the effects of hourly heat appear to be additively separable, motivating this specification.\footnote{Denoting temperature in Celsius as $T_{h}$ for each hour $h$ in growing season year $y$, these two variables are constructed:
\begin{align*}
{degree\_days\_below\_29^\circ C} &= \frac{1}{24}\sum_{h\in y}\left[\max(T_{h}, 0) -\max(T_{h}-29, 0)\right] \\
%$$\mathrm{degree\_days\_below\_29^\circ C} = \frac{1}{24}\sum_{h\in y}\left(\max(T_{h}, 0)\cdot\mathbbm{1}{[T_{h} < 29]} + 29\cdot\mathbbm{1}{[T_{h} \geq 29]}\right) $$
{degree\_days\_above\_29^\circ C} &= \frac{1}{24}\sum_{h\in y}\max(T_{h}-29, 0)
\end{align*}
These variables thus summarize hourly thermal exposure integrated over time across these two thermal ranges in units of \textit{degree-days}.
} %They also show this specification has strong predictive accuracy out of sample. 

Our explanatory variables also include linear terms in the five soil characteristics described above and linear and quadratic terms in total growing-season precipitation (March to August, measured in mm). We note that the SFD approach is well-suited to capture non-linear effects, in this case those of rainfall and temperature, so long as the terms that describe the nonlinearity are computed prior to differencing.\footnote{To see this, note that we construct the SFD estimator for the model $y_i = \beta_1 p_i + \beta_2 p_i^2$ by writing $\Delta y_i= y_i - y_{i-1}  = \beta_1 (p_i  - p_{i-1})+ \beta_2 (p_i^2 -  p_{i-1}^2)=\beta_1 \Delta p_i + \beta_2  \Delta p_i^2$. The coefficient $\beta_2$ maintains its interpretation even after differencing.}

To evaluate the performance of the levels and SFD models in the presence of omitted variables, we initially employ seven different specifications, altering which covariates are included in the regression. Specification (1) includes only temperature variables, (2) includes only rainfall variables, and (3) includes only soil variables. Specification (4) includes temperature and rainfall variables, (5) includes temperature and soil variables, and (6) includes rainfall and soil variables. Specification (7) includes all variables.  For both temperature and precipitation, across all models, the two variables describing each (e.g. \textit{precipitation} and \textit{precipitation-squared}) are either included together or omitted together. By intentionally withholding known covariates from specifications (1)-(6), we mimic situations in which some (possibly unknown) variables are omitted from the model. We then evaluate how the levels and SFD estimators perform in these cases compared to specification (7), the most saturated model.

 \subsection*{Levels estimation}
 
First, we estimate the effect of environmental conditions on maize yields using a standard cross-sectional specification in levels, analogous to the models estimated by \cite{mendelsohn1994impact} and \cite{schlenker2006impact}. The estimate the model 
\beq
log(y_i) = \alpha_1 + \mathbf{t}_i \beta_L + \mathbf{p}_i \gamma_L + \mathbf{s}_i \delta_L + \epsilon_i
\label{Eq:Model_Levels}
\eeq
where $y_i$ is average maize yield in county $i$, $\alpha_1$ is a constant, $\mathbf{t}_i$ is the vector containing \textit{degree-days below 29\degree C} and \textit{degree-days above 29\degree C}, $\mathbf{p}_i$ is the vector containing \textit{precipitation} and \textit{precipitation-squared}, $\mathbf{s}_i$ is the vector of soil variables, and $\epsilon_i$ are unexplained variations. Terms are withheld in specifications (1)-(6) and the full model is estimated in specification (7).  

\begin{table}[t!]
\thispagestyle{empty}
\fontsize{8}{8}\selectfont
\begin{center}
\begin{tabular}{ l c c c c c c c }
\hline
\hline
 & & & & & & & \\
& \multicolumn{7}{c}{\textit{Dependent variable: log maize yield $\times$ 1,000}} \\
 & & & & & & &  \\
\cline{2-8} 
 & & & & & & & \\
 & (1) & (2) & (3) & (4) & (5) & (6) & (7) \\
  & & & & & & & \\
 \hline
 & & & & & & & \\
 Degree-days below 29\degree C  & 0.06 &  &  & $-$0.22 & 0.29$^{**}$ &  & 0.01 \\ 
  & (0.15) &  &  & (0.18) & (0.14) &  & (0.13) \\ 
  & & & & & & & \\ 
 Degree-days above 29\degree C & $-$6.11$^{**}$ &  &  & $-$2.19 & $-$7.98$^{***}$ &  & $-$4.74$^{**}$ \\ 
  & (2.54) &  &  & (2.28) & (2.21) &  & (1.89) \\ 
  & & & & & & & \\ 
 Precipitation &  & 11.45$^{***}$ &  & 10.49$^{***}$ &  & 8.15$^{***}$ & 7.34$^{***}$ \\ 
  &  & (3.84) &  & (2.33) &  & (3.06) & (1.85) \\ 
  & & & & & & & \\ 
 Precipitation-squared &  & $-$0.01$^{***}$ &  & $-$0.01$^{***}$ &  & $-$0.01$^{**}$ & $-$0.01$^{***}$ \\ 
  &  & (0.003) &  & (0.002) &  & (0.003) & (0.002) \\ 
  & & & & & & & \\ 
 Water capacity &  &  & 31.61$^{***}$ &  & 43.19$^{***}$ & 32.81$^{***}$ & 41.21$^{***}$ \\ 
  &  &  & (10.75) &  & (7.36) & (10.66) & (7.95) \\ 
  & & & & & & & \\ 
 Percent clay  &  &  & $-$2.44 &  & $-$0.47 & $-$1.91 & $-$0.07 \\ 
  &  &  & (2.49) &  & (2.21) & (2.41) & (2.01) \\ 
  & & & & & & & \\ 
 Minimum permeability &  &  & 68.68$^{***}$ &  & 69.46$^{***}$ & 61.95$^{***}$ & 57.00$^{***}$ \\ 
  &  &  & (15.94) &  & (18.05) & (17.26) & (16.27) \\ 
  & & & & & & & \\ 
 Soil erodibility factor &  &  & 1,148.55$^{***}$ &  & 816.64$^{***}$ & 719.33$^{*}$ & 477.45$^{*}$ \\ 
  &  &  & (398.16) &  & (234.22) & (408.26) & (254.20) \\ 
  & & & & & & & \\ 
 Best soil class &  &  & 4.59$^{***}$ &  & 3.63$^{***}$ & 3.89$^{***}$ & 2.78$^{***}$ \\ 
  &  &  & (1.19) &  & (0.87) & (1.20) & (0.82) \\ 
  & & & & & & & \\ 
 Constant & 4,487$^{***}$ & 1,394 & 3,432$^{***}$ & 2,110$^{*}$ & 2,838$^{***}$ & 1,287 & 1,531$^{*}$ \\ 
  &  (549) & (1,360) &  (163) & (1,258) &  (565) & (1,013) & (756)  \\ 
  & & & & & & & \\ 
\hline 
 & & & & & & & \\
Observations & 825 & 825 & 825 & 825 & 825 & 825 & 825 \\ 
R squared & 0.29 & 0.28 & 0.39 & 0.47 & 0.58 & 0.50 & 0.66 \\ 
& & & & & & & \\
 \hline
 \hline
 \end{tabular}
 \caption{\textbf{Cross-sectional estimates of the effect of environmental conditions on maize yields using the levels model}. Data are taken from Schlenker and Roberts (2009) and Schlenker, Hanemann, and Fisher (2006) and are for US counties east of the 100th meridian for the years 1950 to 2005. Standard errors account for spatial autocorrelation following Conley (1999). Asterisks indicate statistical significance at the 0.1\%,***, 1\%**, and 5\%* levels.}
 \label{Tab:Levels}
\end{center}
\end{table}

The results are displayed in Table \ref{Tab:Levels}. In the levels research design, the parameter estimates generally have the same sign across models but exhibit highly inconsistent point estimates in almost all cases. For example, the estimated effects of moderate temperature ({degree-days below 29\degree C}) and extreme heat ({degree-days above 29\degree C}) change substantially between the specification where only temperature is included (1) and the specification where both temperature and precipitation are included (4). Indeed, the coefficient estimate for days with moderate temperatures is positive in (1) but negative in (4), and the estimate for extreme heat is three times larger in (1) than it is in (4). The estimated effects for the soil variables also change considerably across specifications. For instance, across the five soil characteristics, the average difference between the estimates in the specification with only soil controls (3) and with all variables (7) is 50\% of the point estimate in specification (3), with a high of 97\% (for percent clay) and a low of 17\% (for minimum permeability). Finally, we find it is worrisome that the estimated coefficient for the soil erodibility factor is positive and significant in all models, at odds with the agronomic literature that finds higher soil erodibility generally decreases yields \citep{renard1997predicting}. 

These results highlight the vulnerability of the standard levels model to omitted variables bias. Indeed, withholding covariates generally leads to large changes in the magnitude of estimated effects. In the case of degree-days below 29\degree C the sign of estimates are inconsistent across models, and in the case of the soil erodibility the estimate has the incorrect sign and is ``statistically significant" across all models.  Even when we include all three sets of controls in specification (7), we cannot be certain that important variables are not still missing, and given the inconsistency of parameter estimates across observed specifications, it is not unreasonable to expect that the estimates may once again change substantially if we included additional covariates in the regression model \citep[e.g.][]{oster2014unobservable}.
 
\subsection*{Spatial First Differences estimation}

Next, we estimate the effect of environmental conditions on maize yields using the SFD research design. To do this, we first must overcome a key challenge to implementing SFD with county-level data: administrative boundaries do not follow the regular lattice structure depicted in Figure \ref{Fig:Implementation}b. There exist many adjacencies between counties that may be exploited in an SFD research design, but ensuring that each county is differenced from exactly one neighboring county in a sequence is no longer trivial. Equation (\ref{Eq:SFDEstimator}) does not define a specific way in which to define neighbors and many arrangements would be valid. Importantly, sampling of differences must be organized such that no observation is double counted. Here, we develop a generalizable approach that imposes a ``panel-like'' structure on irregularly shaped counties, thereby identifying sequences of adjacent counties without double counting them. Notably, however, there exist other valid algorithms for setting up SFD in two dimensional space that we do not explore here.

 \begin{figure}[t]
\begin{center}
\includegraphics[width=\textwidth]{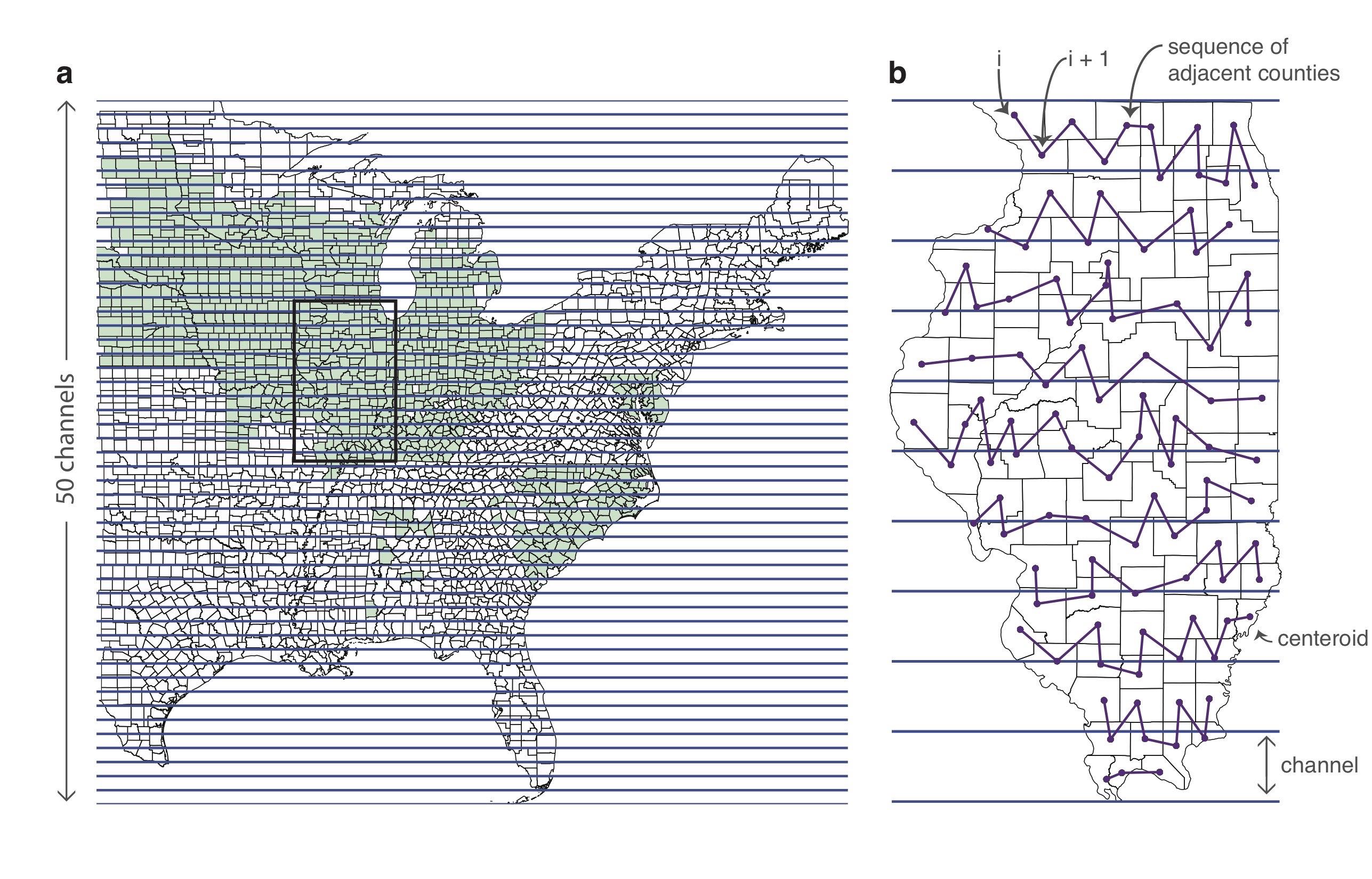}
\caption{{\bf Sampling procedure for spatial first differences with irregular (non-gridded) data.} (a) US counties east of the 100th meridian with sampling channels overlaid in blue. Counties included in the balanced panel are highlighted shaded green. (b) Detail of insert in a, depicting the algorithm used to generate sequences of adjacent counties to construct a ``panel-like" data structure, analogous to Figure \ref{Fig:Implementation}b (see text for description).}
\label{Fig:SamplingProcedure}
\end{center}
\end{figure}

The basic procedure is depicted in Figure \ref{Fig:SamplingProcedure}. First, we overlay the US map with 50 sampling ``channels." These are long and narrow regions, approximately 30 miles wide, spanning West-East slices of the country and defined by a northern and southern boundary.\footnote{The width of our sampling channels was chosen to match the average width (from North to South) of the counties in our sample.} Adjacent channels share a boundary. Beginning with the northernmost channel, all the counties that intersect with the channel are recorded as sequentially adjacent and ordered by the longitude of their centroid. Then we move south to the next channel and repeat this process. To assure that each county is only included in one channel, if a county has already been sampled in a preceding more northern channel, it is omitted from the remaining southern channels. The sequence of counties within each channel are thus treated like a sequence of observations within a ``panel-like unit'' of the regularly shaped observations shown in Figure \ref{Fig:Implementation}b.  Finally, differences are computed between the ordered adjacent counties and a cross-sectional regression is estimated in these first-differences. Notably, SFD can be implemented in any direction by rearranging how the channels are oriented when they are first overlaid. We begin our analysis by computing SFD in both the West-East direction and in the North-South direction for comparison.

\begin{figure}
\vspace*{-0.5cm}
\begin{center}
\includegraphics[width = .6\textwidth]{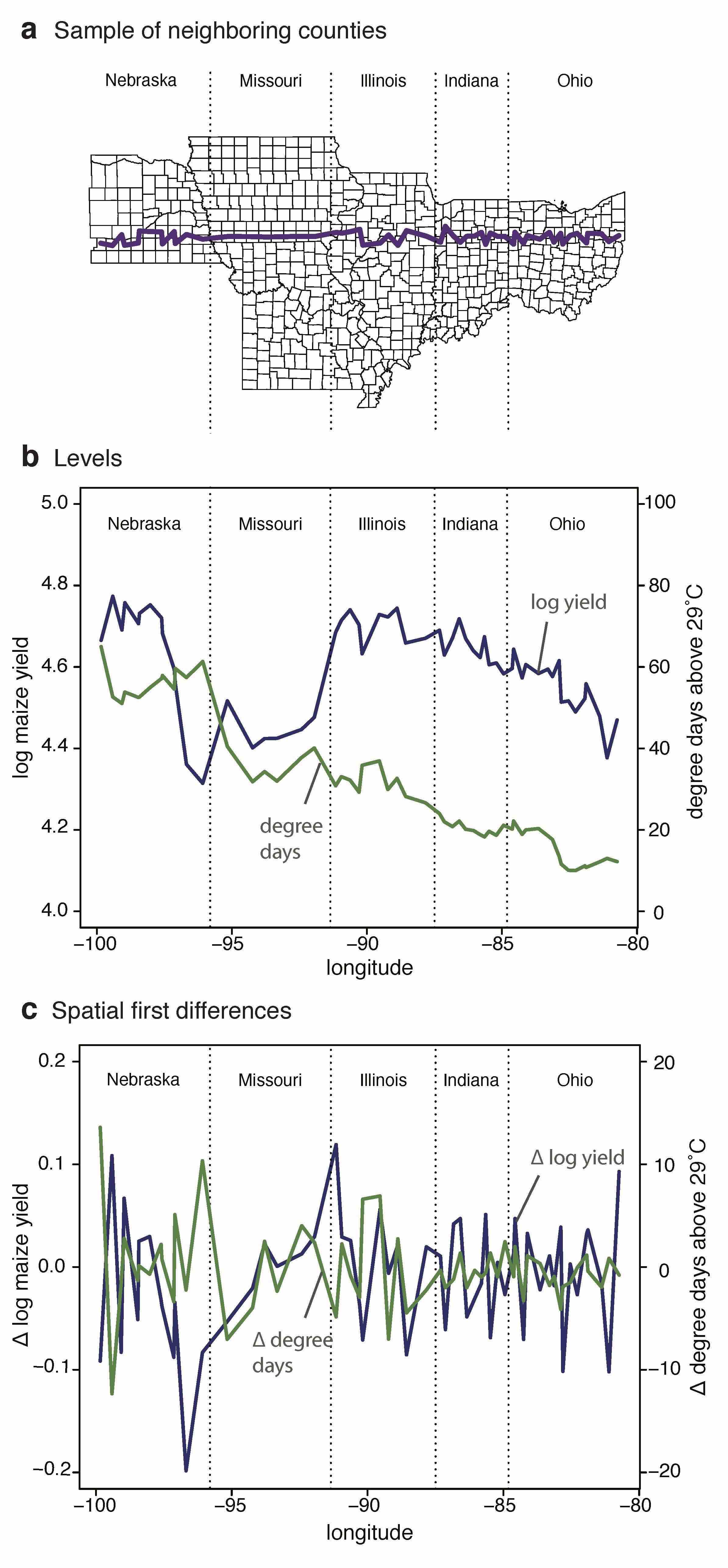}
\caption{{\bf Comparison of levels and SFD in agricultural data.} (a) The sequence of adjacent counties included in a single sampling ``channel" (purple). (b) Log annual maize yield (blue) and number of days with temperatures above 29\degree C (green). (c) Shows the same data as in panel b after taking spatial first differences.}
\label{Fig:LevelsVDiffs}
\end{center}
\end{figure}

Figure \ref{Fig:LevelsVDiffs} illustrates a single sequence of adjacent counties within one channel (panel a) derived using this procedure and compares two variables of interest in levels and SFD. Comparing levels (panel b) to SFD (panel c), one can see how the use of spatial first differences eliminates low-frequency correlations contained in the ``spatial history" of the two variables. What remains is the variation we use to estimate $\hat{\beta}_{SFD}$. Note that discontinuities in the levels of these variables, such as those that occur at the borders of Missouri, do not systematically affect the data after differencing. 

Using the differences computed between ordered adjacent neighbors, we estimate the effect of climate and soil on maize yields via the SFD model
\beq
\Delta log(y_i) = \alpha_2 + \Delta \mathbf{t}_i \beta_{SFD} + \Delta \mathbf{p}_i \gamma_{SFD} + \Delta \mathbf{s}_i \delta_{SFD} + \Delta \epsilon_i. 
\label{Eq:Model_SFD}
\eeq
The model now contains differences of terms in non-linear functions for temperature and precipitation, but the coefficients for these differenced terms maintain the same interpretation as in the levels model.

The results for the same set of specifications (1)-(7) are displayed in Table \ref{Tab:SFD}, although each specification is now estimated twice, once in the West-East direction and once in the North-South direction. Using SFD, the coefficient estimates are extremely consistent across models, especially in comparison to the levels estimates in Table \ref{Tab:Levels}. There are no sign reversals and the median difference between the coefficient estimate in the model with just one set of variables and the model with all three sets of variables is 12\% of the point estimate in the model with one set, as opposed to 38\% for the levels estimates.  Additionally, the SFD estimates are essentially unchanged when calculated in the West-East and North-South directions. The median difference between the West-East estimate and the North-South estimate is 10\% of the point estimate, and this difference is less than 30\% for all but one variable (percent clay).

%\begin{landscape}
\begin{sidewaystable}
\begin{adjustwidth}{-2cm}{}
\fontsize{8}{8}\selectfont
\begin{center}
\begin{tabular}{ l c c c c c c c c c c c c c c }
\hline
\hline
 & & & & & & & & & & & & & & \\
& \multicolumn{14}{c}{\textit{Dependent variable: average log maize yield $\times$ 1,000}} \\
 & & & & & & & & & & & & & & \\
 & (1 WE) & (1 NS) & (2 WE) & (2 NS) & (3 WE) & (3 NS) & (4 WE) & (4 NS) & (5 WE) & (5 NS) & (6 WE) & (6 NS) & (7 WE) & (7 NS) \\
  & & & & & & & & & & & & & & \\
 \hline
 & & & & & & & & & & & & & & \\
 
 Degree-days below 29\degree C & 0.35$^{***}$ & 0.43$^{***}$ &  &  &  &  & 0.32$^{**}$ & 0.38$^{***}$ & 0.31$^{***}$ & 0.38$^{***}$ &  &  & 0.28$^{**}$ & 0.33$^{***}$ \\ 
  & (0.13) & (0.12) &  &  &  &  & (0.13) & (0.11) & (0.11) & (0.12) &  &  & (0.11) & (0.11) \\ 
  & & & & & & & & & & & & & & \\ 
 Degree-days above 29\degree C  & $-$4.99$^{***}$ & $-$4.77$^{***}$ &  &  &  &  & $-$4.71$^{**}$ & $-$4.29$^{***}$ & $-$5.12$^{***}$ & $-$4.93$^{***}$ &  &  & $-$4.73$^{***}$ & $-$4.41$^{***}$ \\ 
  & (1.89) & (1.72) &  &  &  &  & (1.96) & (1.65) & (1.59) & (1.52) &  &  & (1.71) & (1.49) \\ 
  & & & & & & & & & & & & & & \\ 
 Precipitation &  &  & 3.72$^{**}$ & 4.10$^{***}$ &  &  & 3.33$^{**}$ & 3.61$^{***}$ &  &  & 2.98$^{**}$ & 3.44$^{***}$ & 2.52$^{**}$ & 2.99$^{***}$ \\ 
  &  &  & (1.48) & (1.13) &  &  & (1.39) & (1.15) &  &  & (1.33) & (0.89) & (1.22) & (0.92) \\ 
  & & & & & & & & & & & & & & \\ 
 Precipitation-squared &  &  & $-$0.003$^{**}$ & $-$0.003$^{***}$ &  &  & $-$0.003$^{**}$ & $-$0.003$^{***}$ &  &  & $-$0.002$^{*}$ & $-$0.002$^{***}$ & $-$0.002$^{**}$ & $-$0.002$^{***}$ \\ 
  &  &  & (0.001) & (0.001) &  &  & (0.001) & (0.001) &  &  & (0.001) & (0.001) & (0.001) & (0.001) \\ 
  & & & & & & & & & & & & & & \\ 
 Water capacity &  &  &  &  & 19.82$^{***}$ & 19.26$^{***}$ &  &  & 17.87$^{***}$ & 17.31$^{***}$ & 20.02$^{***}$ & 18.55$^{***}$ & 18.37$^{***}$ & 17.09$^{***}$ \\ 
  &  &  &  &  & (4.53) & (4.65) &  &  & (4.67) & (4.19) & (4.56) & (4.58) & (4.67) & (4.24) \\ 
  & & & & & & & & & & & & & & \\ 
 Percent clay &  &  &  & & $-$0.43 & $-$2.01$^{**}$ &  &  & $-$0.31 & $-$1.89$^{**}$ & $-$0.16 & $-$1.66 & $-$0.13 & $-$1.67$^{*}$ \\ 
  &  &  &  &  & (1.04) & (1.02) &  &  & (0.96) & (0.94) & (1.09) & (1.04) & (0.99) & (0.96) \\
  & & & & & & & & & & & & & & \\ 
 Minimum permeability &  &  &  & & 16.67$^{**}$ & 14.01 &  &  & 17.91$^{***}$ & 15.07 & 16.54$^{**}$ & 14.23 & 17.65$^{***}$ & 15.00$^{*}$ \\ 
  &  &  &  &  & (6.79) & (9.58) &  &  & (6.22) & (9.23) & (6.77) & (9.37) & (6.17) & (9.09) \\ 
  & & & & & & & & & & & & & & \\ 
 Soil erodibility factor &  &  &  &  & $-$348.43$^{*}$ & $-$286.06 &  &  & $-$319.96$^{*}$ & $-$266.26 & $-$363.80$^{*}$ & $-$292.48 & $-$335.54$^{*}$ & $-$278.18 \\ 
  &  &  &  &  & (201.30) & (187.00) &  &  & (178.98) & (178.82) & (203.78) & (186.87) & (179.74) & (179.99) \\ 
  & & & & & & & & & & & & & & \\ 
 Best soil class &  &  &  &  & 2.32$^{***}$ & 2.19$^{***}$ &  &  & 2.49$^{***}$ & 2.31$^{***}$ & 2.28$^{***}$ & 2.20$^{***}$ & 2.43$^{***}$ & 2.29$^{***}$ \\ 
  &  &  &  &  & (0.47) & (0.65) &  &  & (0.41) & (0.55) & (0.46) & (0.60) & (0.41) & (0.54) \\ 
  & & & & & & & & & & & & & & \\ 
 Constant & 4.99 & $-$4.50 & 6.45$^{*}$ & $-$3.47 & 9.75$^{**}$ & 2.41 & 4.11 & $-$6.90$^{**}$ & 6.37$^{*}$ & $-$1.04 & 8.19$^{***}$ & $-$1.74 & 5.61$^{***}$ & $-$3.46 \\ 
  & (3.44) & (2.77) & (3.37) & (2.56) & (3.28) & (2.17) & (3.21) & (3.00) & (2.93) & (2.55) & (1.91) & (2.52) & (1.91) & (2.97) \\ 
  & & & & & & & & & & & & & & \\   
 \hline
  & & & & & & & & & & & & & & \\
 Observations & 804 & 825 & 804 & 825 & 804 & 825 & 804 & 825 & 804 & 825 & 804 & 825 & 804 & 825 \\ 
R squared & 0.04 & 0.04 & 0.02 & 0.03 & 0.24 & 0.20 & 0.05 & 0.06 & 0.28 & 0.24 & 0.25 & 0.22 & 0.28 & 0.25 \\ 
  & & & & & & & & & & & & & & \\
 \hline
 \hline
 \end{tabular}
  \caption{\textbf{SFD estimates of the effect of climate and soil on maize yields}. Data are for US counties east of the 100th meridian with complete time series for the years 1950 to 2005. All variables are averaged over the 56-year time period. SFD estimates are computed both in the West-East (WE) and North-South (NS) directions. Standard errors account for spatial autocorrelation following \cite{conley1999gmm}. Asterisks indicate statistical significance at the 0.1\%,***, 1\%**, and 5\%* levels.}
 \label{Tab:SFD}
\end{center}
\end{adjustwidth}
\end{sidewaystable}
%\end{landscape} 

The SFD estimates are also all consistent with the agronomic literature. Days with moderate temperatures are estimated to significantly increase yields across all specifications, in contrast to the levels model which recovered this result in one of four specifications. As expected, days with extreme heat reduce yields across all specifications. Precipitation changes have a smaller impact, but continue to have an inverted U-shaped effect on yields. A higher average water capacity increases yields, a higher percentage of clay reduces yields, a lower minimum permeability (which indicates drainage problems) reduces yields, a higher soil erodibility factor is harmful (the levels model consistently indicated the opposite), and better soils (as measured by best soil class) are beneficial.  The SFD research design was able to recover these results in all specifications, in both the West-East and North-South directions. The relative invariance of all coefficient estimates across all models provides us with modest confidence that our results are robust to other important covariates that may not be observed even in the most saturated specification.

Our estimates for the constant term, $\hat{\alpha}_2$ are generally positive and significant in the West-East model and have a negative sign in the North-South model. In the SFD model, the constant term is interpretable as average trend in space as one moves in the direction of differencing, conditional on changes in the covariates. Thus a positive constant term in the West-East model indicates that yields are increasing as one moves from West to East (since we have subtracted from each observation values from its neighbor to the West). Similarly, a negative constant term in the North-South model indicates that yields are decreasing as one moves from North to South.

One practical question that arises with empirical estimation is how to calculate standard errors for SFD estimates. In all cases, we expect residuals estimated via OLS to be negatively auto-correlated (at least first-order) due to the first-differencing procedure since sequential residuals $\Delta \epsilon_i = \epsilon_i - \epsilon_{i-1}$ and $\Delta \epsilon_{i+1} = \epsilon_{i+1} - \epsilon_{i}$ share the component $\epsilon_i$ and it enters positively in one instance and negatively in the other. In this context, we also reject the null hypothesis of homoskedastic disturbances using the Bruesh-Pagan test. To address these two issues, we calculate five different sets of standard errors: (i) Conley standard errors, (ii) Newey-West standard errors, (iii) standard errors clustered by channel (iv) bootstrapped standard errors,\footnote{For the bootstrapped standard errors, we resample at the observation-level after differencing between adjacent counties.} and (v) block-bootstrapped standard errors block-resampled by sampling channel. These standard errors are presented in Appendix \ref{AppendixSE}, along with the OLS standard errors for comparison. The magnitude of these various standard error estimates are comparable across all five methods, but the Conley, Newey-West, and block bootstrapped standard errors are generally slightly larger, suggesting some additional spatial autocorrelation in $\Delta \epsilon$ beyond immediate neighbors. Thus, in Table \ref{Tab:SFD}, we report standard errors that account for spatial autocorrelation using Conley's approach.

\begin{figure}[t!]
\begin{center}
\includegraphics[width = \textwidth]{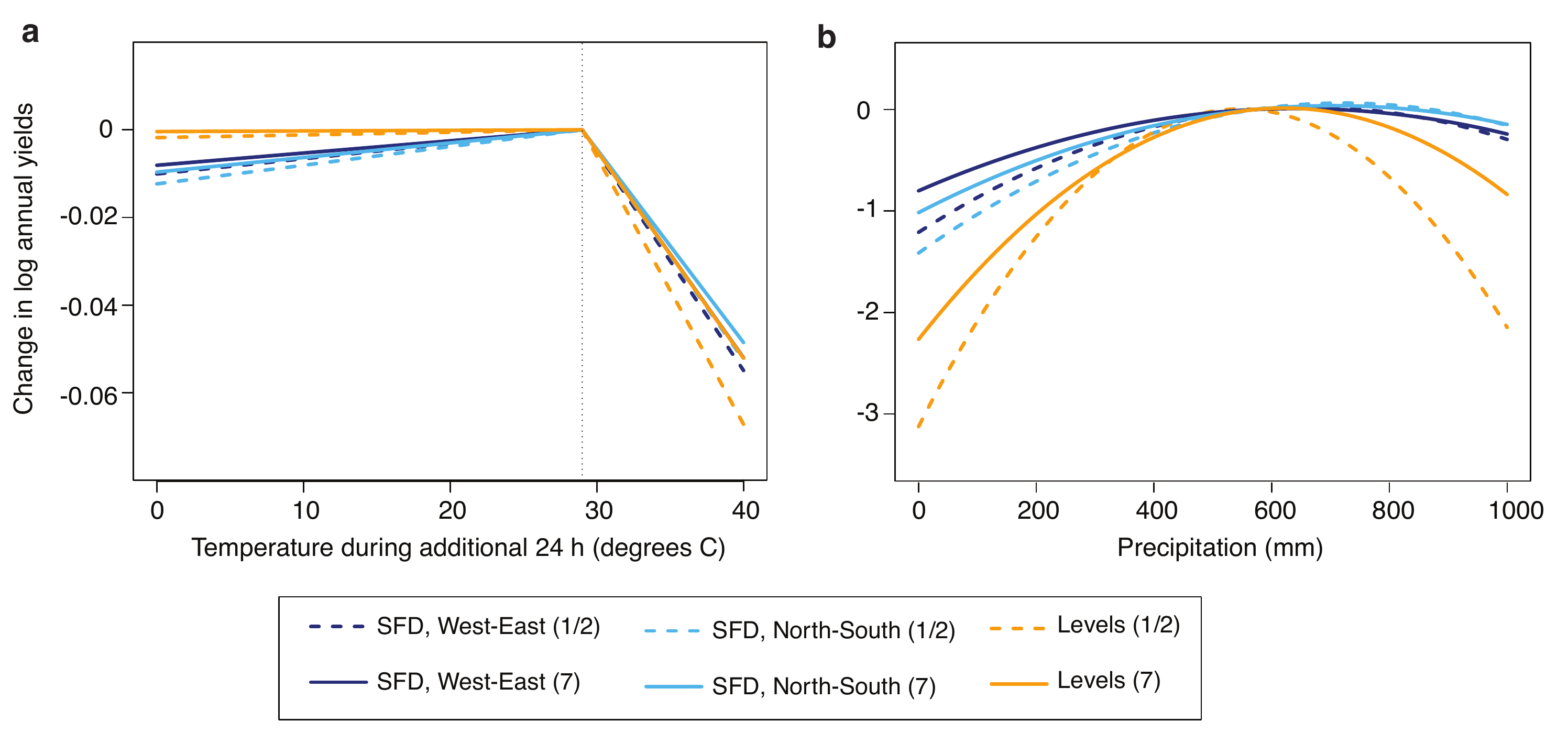}
\caption{{\bf Estimated effects of climate on maize yield in the US cross section.} (a) Comparison of the effect of daily temperature on maize yields estimated using levels (orange) and SFD (blue). (b)  Same but for precipitation. Specifications are denoted in parentheses in the legend. Variables are 56 year averages.}
\label{Fig:TempPlot}
\end{center}
\end{figure}

The recent literature in this field has been particularly interested in the effect of climate on yields, motivated by efforts to understand the economic consequences of climate change \citep[e.g.][]{mendelsohn1994impact, deschenes2007economic, schlenker2009nonlinear, burke2016adaptation, hsiang2016climate}. Because the temperature-yield and precipitation-yield relationships are nonlinear and the coefficient estimates are difficult to interpret, we plot these relationships in Figure  \ref{Fig:TempPlot}. Results from the levels model are shown in orange and those from the SFD model are shown in blue. We include the estimates from both the specification with no controls ((1) for temperature, (2) for precipitation) and the model with a full set of controls (7). Panel a shows our estimated effects for temperature. Two features of the results stand out. First, the two SFD estimates are very near one another, despite the differences being computed in orthogonal directions and thus exploiting different variation in the independent variables. Second, the SFD estimates are remarkably similar to previous estimates of the effect of long-run trends in temperature on US maize yields. Our coefficient estimate for the variable \textit{degree-days above 29\degree C} is $-0.0050$ when SFD are computed in the West-East direction and $-0.0048$ when SFD are computed in the North-South direction. Using long differences over the period 1980-2000, \cite{burke2016adaptation} estimated this same coefficient to be $-0.0053$ in their specification with time-specific fixed effects and $-0.0044$ in their specification with state fixed effects.\footnote{We use \cite{burke2016adaptation} as a benchmark against which to compare our results because it is the only study to estimate the effect of long-run climate on agricultural productivity that is plausibly robust to unobserved heterogeneity. The authors employ a ``long differences" approach and model county-level changes in yields over time as a function of changes in temperature and precipitation, accounting for time-invariant unobservables at the county level and time-trending unobservables at the state level.} Holding all else equal, these findings imply\footnote{$-0.00499$ per degree day $\times10$ degree-days $= -0.05$, i.e. 5 log points.} that substituting one full day (24 hours) at 29\degree C temperature with a full day at 40\degree C results in a predicted end-of-season yield decline of approximately 5\%. Panel b of Figure \ref{Fig:TempPlot} shows the estimated effect of precipitation on maize yields. Once again, the SFD estimates are near one another across different regression models and differencing directions. Also noteworthy is the result that the four different SFD estimates largely agree with one another while the two levels estimates differ substantively, both from one another and from the range of SFD estimates. 

\subsection*{Systematic omission of variables}

With the goal of systematically evaluating both research designs' vulnerability to omitted variables bias, we experiment further by withholding all possible combinations of covariates when estimating effects for each explanatory variable. For each environmental variable of interest (e.g. temperature), this procedure generates a total of 192 different specifications where the remaining six controls are systematically withheld in all possible combinations (e.g. precipitation, precipitation $+$ water capacity, precipitation $+$ water capacity $+$ percent clay, $\dots$).\footnote{The temperature variables (\textit{degree-days below 29\degree C} and \textit{degree-days above 29\degree C}) always appear together. Similarly, the \textit{precipitation-squared} variable is only included when \textit{precipitation} is included.} This type of analysis is similar to the ``extreme bounds" analysis proposed by \cite{leamer1985sensitivity}, and taken to its logical extreme by \cite{sala1997just}, who ran nearly two million growth regressions using different combinations of 62 explanatory variables. The goal of the procedure, as we are using it, is to gain more general insight into the magnitude and distribution of the omitted variables bias that is eliminated from the cross-sectional levels regression by differencing, as described in Eq. (\ref{Eq:OVBComparison}). Specifically, for each variable of interest, we compute all possible estimates for $\hat{\beta}_L$ and $\hat{\beta}_{SFD}$ and compare their relative stability across these specifications. Variations in $\hat{\beta}$ that occur when covariates change are interpreted as evidence of omitted variables bias, although it is unknown which specification is ``correct." 

\begin{figure}[t!]
\begin{center}
\vspace{-9mm}
\includegraphics[width = \textwidth]{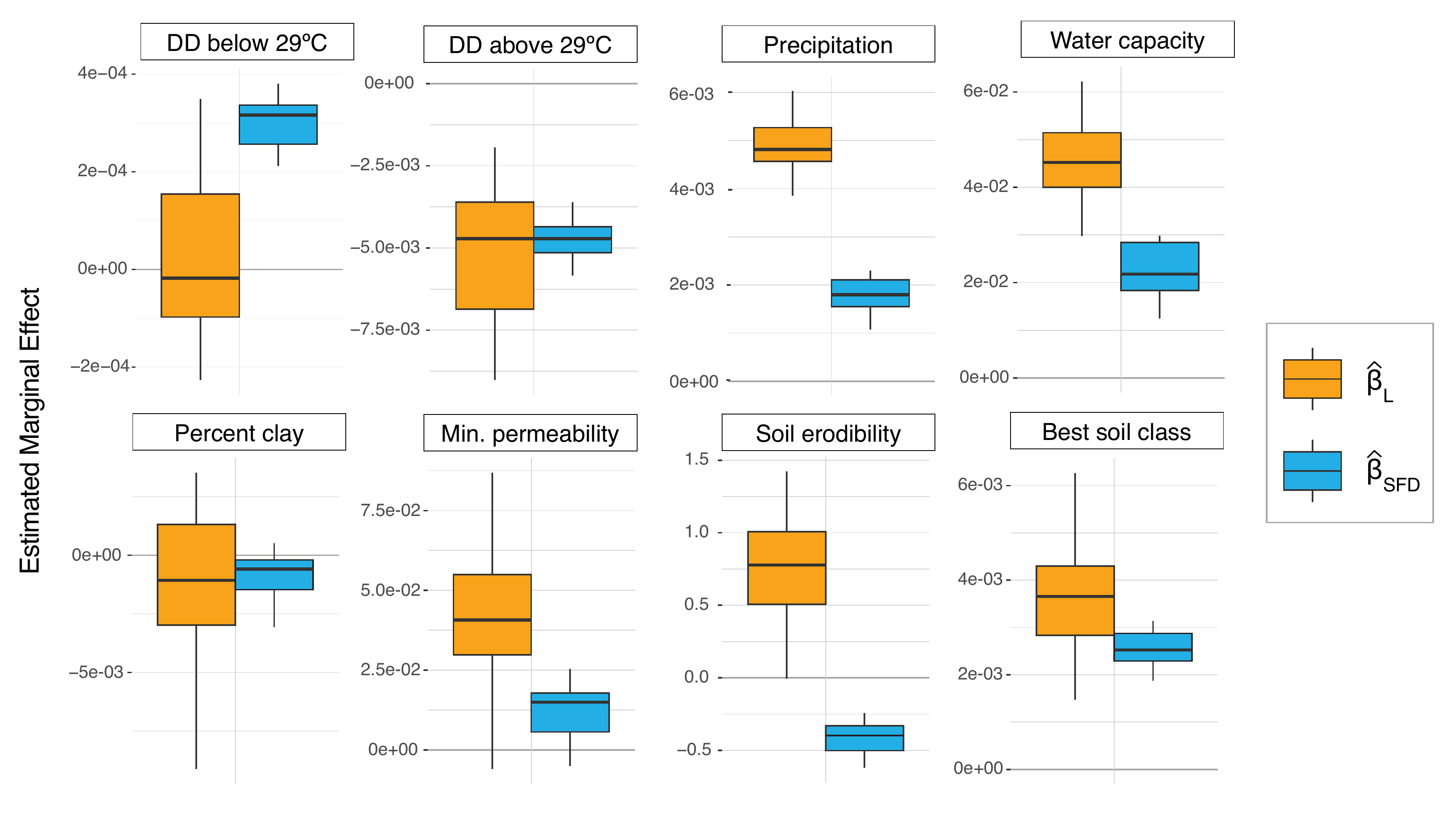}
\caption{{\bf Marginal effect estimates across all combinations of covariates.} The distributions of estimated marginal effects obtained for each variable across 192 models that contain all possible sets of the six remaining variables as covariates.  Boxes show the interquartile range of estimates and whiskers show the maximum and minimum estimates. Regressions in levels are orange, SFD are blue. The two degree-day measures (``DD'') are always included or excluded together.  The displayed marginal effect of precipitation is calculated at the median (572 mm). }
\label{Fig:Boxplots}
\end{center}
\end{figure}

Using this extreme bounds analysis, SFD dramatically outperforms estimation in levels. The distribution of estimated marginal effects across the 192 regression specifications for each variable is shown in Figure \ref{Fig:Boxplots}. The variance of this distribution (averaged across variables) is 80\% smaller when employing SFD as opposed to a cross section on levels, with a high of 100\% (for best soil class) and a low of 47\% (for water capacity). Furthermore, under the SFD model, the coefficients for days with moderate heat and the soil erodibility factor have the expected positive and negative signs, respectively, across all 192 specifications. This pattern does not hold for the levels model, where the coefficient for days with moderate heat is often negative and the coefficient for the soil erodibility factor always has the wrong sign (positive). Reinforcing our findings from above, these results suggest that including/omitting variables often leads to substantial changes in the levels estimates but has limited effect on SFD estimates in this context.\\

This example demonstrates how the SFD research design can be robust to unobserved heterogeneity. In the context of maize yields, there appears to be a large degree of low-frequency spatial correlation between the covariates, leading to large biases in $\hat{\beta}_L$ when control variables are withheld from the model. In contrast, SFD recovers estimates for all seven environmental factors that are essentially unchanged regardless of whether or not key variables are included in the model. We suspect that if an important covariate were still missing from the saturated regression model (7) and it was discovered and included in a new specification, it would not dramatically change the SFD estimates.

\subsection*{Novel robustness checks unique to Spatial First Differences}

As a final exercise, we demonstrate two internal robustness checks made uniquely possible in the SFD research design: a rotation of the coordinate system and spatial double differences (SDD). As discussed above, these tests should fail if the orthogonality condition in Eq. (\ref{Eq:IndepCondition}) is not true. 

\begin{figure}[t]
\begin{center}
\vspace{-10mm}
\includegraphics[width = .8\textwidth]{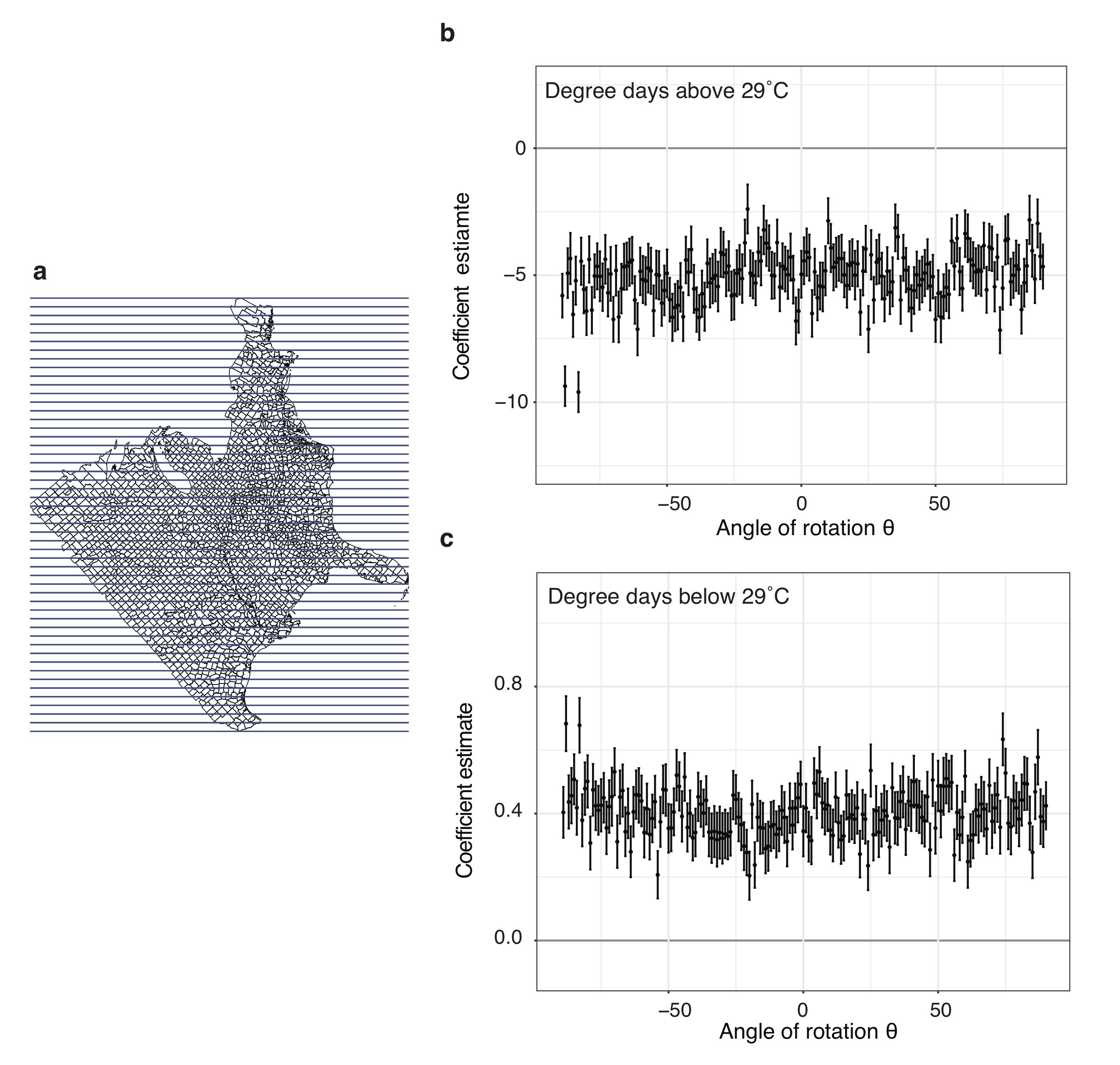}
\caption{{\bf Rotation of the coordinate system.} (a) Demonstrates the $\theta = 45\degree$ rotation of the coordinate system for counties relative to the sampling channels. (b) estimated marginal effects and standard errors for extreme heat (degree-days above 29\degree C) using SFD at each angle of rotation $\theta$ (from -89\degree to 90\degree). (C) Same, but for moderate temperatures (degree-days below 29\degree C).}
\label{Fig:Rotation}
\end{center}
\end{figure}
First, we conduct a sensitivity analysis that exploits the rotation of the coordinate system. After imposing the sampling grid on the US map with the channels arranged in the West-East direction, we rotate the map by an angle $\theta$ at 1\degree increments on its axis under the grid for $\theta = -89\degree$ to $90\degree$ (see Figure \ref{Fig:Rotation}a).\footnote{Equivalently, one could instead choose to keep the position of the map fixed, and rotate the sampling channels by the angle $-\theta$.}  We then estimate the nonlinear effect of temperature on maize yields using SFD with no controls at each $\theta$, producing 180 different estimates of specification (1). These 180 estimated marginal effects of extreme heat and moderate temperatures are shown in Figures \ref{Fig:Rotation}b and \ref{Fig:Rotation}c, respectively. Note that the estimated effects are highly consistent across the different coordinate rotations. Indeed, the variance of the coefficient estimate for \textit{degree-days above 29\degree C} is $1.025$ for a coefficient estimate of $-5$, producing a coefficient of variation equal to 0.2 (Figure \ref{Fig:Rotation}b). Similarly, \textit{degree-days below 29\degree C} have a coefficient of variation equal to 0.19 (Figure \ref{Fig:Rotation}c). The fact that the estimates are consistent across all sampling directions implies either the identifying assumption holds (i.e. Eq. \ref{Eq:IndepCondition} is true) or that it fails but somehow generates a similar bias for each of the 180 different estimates, despite these estimates exploiting different sources of variation in the independent variables.

\begin{table}[t]
\fontsize{8}{8}\selectfont
\begin{center}
\begin{tabular}{ l c c c c c c c c }
\hline
\hline
 & & & & & & & &  \\
& \multicolumn{8}{c}{\textit{Dependent variable: average log maize yield $\times$ 1,000}} \\
 & & & & & & & & \\
\cline{2-9} 
 & & & & & & & &  \\
 & (1 WE) & (1 NS) & (2 WE) & (2 NS) & (3 WE) & (3 NS) & (7 WE) & (7 NS) \\
 & & & & & & & &  \\
 \hline
 & & & & & & & &  \\
 Degree-days below 29\degree C  & 0.17 & 0.40$^{**}$ &  &  &  &  & 0.16 & 0.26$^{**}$ \\ 
  & (0.16) & (0.16) &  &  &  &  & (0.13) & (0.13) \\ 
  & & & & & & & & \\ 
 Degree-days above 29\degree C & $-$3.47 & $-$4.16$^{**}$ &  &  &  &  & $-$3.65$^{*}$ & $-$3.17$^{*}$ \\ 
  & (2.18) & (2.07) &  &  &  &  & (1.94) & (1.85) \\ 
  & & & & & & & & \\ 
 Precipitation C&  &  & 3.21$^{**}$ & 4.84$^{***}$ &  &  & 2.68$^{**}$ & 3.29$^{*}$ \\ 
  &  &  & (1.42) & (1.85) &  &  & (1.33) & (1.71) \\ 
  & & & & & & & & \\ 
 Precipitation-squared  &  &  & $-$0.002$^{**}$ & $-$0.003$^{**}$ &  &  & $-$0.002$^{*}$ & $-$0.002 \\ 
  &  &  & (0.001) & (0.002) &  &  & (0.001) & (0.001) \\ 
  & & & & & & & & \\ 
 Water Capacity  &  &  &  & & 19.24$^{***}$ & 21.27$^{***}$ & 18.94$^{***}$ & 19.29$^{***}$ \\ 
  &  &  &  &  & (4.84) & (4.41) & (4.94) & (4.23) \\
  & & & & & & & & \\ 
 Percent Clay &  &  &  &  & 0.28 & $-$3.84$^{***}$ & 0.54 & $-$3.42$^{***}$ \\ 
  &  &  &  &  & (1.12) & (1.35) & (1.02) & (1.16) \\ 
  & & & & & & & & \\ 
 Minimum permeability  &  &  &  &  & 21.96$^{***}$ & 7.04 & 22.58$^{***}$ & 8.93 \\ 
  &  &  &  &  & (6.50) & (7.62) & (5.98) & (6.87) \\ 
  & & & & & & & & \\ 
 Soil erodibility factor &  &  &  &  & $-$345.06$^{*}$ & $-$386.77$^{*}$ & $-$359.54$^{**}$ & $-$355.49$^{*}$ \\ 
  &  &  &  &  & (192.81) & (232.53) & (182.15) & (207.49) \\ 
  & & & & & & & & \\ 
 Best soil class &  &  &  &   & 2.36$^{***}$ & 1.79$^{***}$ & 2.38$^{***}$ & 1.87$^{***}$ \\ 
  &  &  &  &  & (0.44) & (0.38) & (0.39) & (0.34) \\ 
  & & & & & & & & \\ 
 Constant & $-$3.85$^{**}$ & $-$6.06$^{***}$ & $-$3.25 & $-$6.50$^{***}$ & $-$2.67$^{*}$ & $-$5.52$^{***}$ & $-$2.59$^{**}$ & $-$5.00$^{***}$ \\ 
  & (1.80) & (1.80) & (2.26) & (2.26) & (1.52) & (1.52) & (1.31) & (1.31) \\ 
  & & & & & & & & \\  \hline
 & & & & & & & &  \\
Observations & 737 & 753 & 737 & 753 & 737 & 753 & 737 & 753 \\ 
R squared & 0.02 & 0.03 & 0.01 & 0.04 & 0.27 & 0.21 & 0.30 & 0.25 \\ 
 & & & & & & & &  \\
 \hline
 \hline
 \end{tabular} 
 \caption{\textbf{SDD estimates of the effect of environmental conditions on maize yields}. Data are for US counties east of the 100th meridian for the years 1950 to 2005. SDD estimates are computed both in the West-East (WE) and North-South (NS) directions. Standard errors account for spatial autocorrelation following Conley (1999).  Asterisks indicate statistical significance at the 0.1\%,***, 1\%**, and 5\%* levels.}
\label{Tab:SDD}
\end{center}
\end{table}

Second, we repeat the analysis using SDD, as described in Eq. (\ref{Eq:SDDModel}). The SDD estimates are displayed in Table \ref{Tab:SDD}. These estimates are near the SFD estimates (with the exception of those for \textit{percent clay}), with a median difference of 18\% of the SFD coefficient estimate. While the percent difference between the SFD and SDD estimates for \textit{percent clay} are large, the SDD estimates for this variable still lie within the 95\% confidence interval of the SFD estimates. In this context, it is not surprising that the SFD estimates and SDD estimates differ somewhat since it is difficult to estimate $\hat{\beta}_{SDD}$ precisely with this modestly sized sample. $\hat{\beta}_{SDD}$ will almost certainly be a substantially more variable estimate than $\hat{\beta}_{SFD}$ in almost all environments as variation in $\Delta^2 \mathbf{x}$ is much smaller than variation in $\Delta \mathbf{x}$. In practice, SDD is also vulnerable to attenuation bias since a relatively larger fraction of variation in $\Delta^2 \mathbf{x}$ may be due to measurement error. Nonetheless, similar to the rotation test above, the stability of estimates across SFD and SDD suggests that in order for $\hat{\beta}_{SFD}$ to be biased by the failure of the identifying orthogonality condition (Eq. \ref{Eq:IndepCondition}), the structure of the omitted variables must be such that $\hat{\beta}_{SDD}$ is similarly biased by the failure of an entirely different orthogonality condition (Eq. \ref{Eq:DFDIndependance}) for each variable. 

We interpret the results of these robustness checks as strong evidence that the assumptions underlying the SFD research design are very likely to be valid in the context of these new empirical estimates.

%-------------------------------------DISCUSSION

\section*{Conclusions}

In standard cross-sectional approaches to inference, it is well understood that the omission of unobservable covariates may lead to large biases in estimated effects. Due to this fact, cross-sectional approaches are often not considered reliable research designs for obtaining causal estimates in many disciplines when instrumental variables are unavailable. We propose SFD as a simple, general, and robust alternative when observations are organized and densely packed in space.  

We highlight that the Local Conditional Independence assumption underlying SFD is conceptually similar to the assumptions exploited in several well-established research designs. These include the assumption that immediately sequential observations within a time series are comparable in event study designs, the assumption that sequential observations within a panel unit are comparable in differences-in-differences panel analyses, and the assumption that observations just above and just below a treatment discontinuity are comparable in regression-discontinuity designs. Indeed, the assumptions necessary for the SFD approach to be valid are so nearly identical to these other assumptions that it seems difficult to logically reject one without also rejecting the other. Importantly, however, SFD is not suitable for all contexts and requires judgement from the analyst about whether observational units are ``dense enough" in space, a data constraint that informs the potential validity of the Local Conditional Independence assumption in practice. 

We imagine that the application of SFD could be applied in a number of different geometries. We demonstrate the application of SFD in one-dimensional space, in two-dimensional gridded data, and in US counties by imposing a ``panel-like" structure on the data. However, with irregular (non-gridded) data, other approaches could be taken. For example, in the two-dimensional space, one could difference in a spiral structure to generate a single sequence of adjacent observations (rather than the ``channels" approach we explore here). One could also combine differences taken in both the West-East and North-South directions---which exploit different variation in the variables---to increase the amount of variation in the sample. The SFD design could also be applied in other contexts. For instance, one might implement SFD along a coastline, throughout an infrastructure network, or even vertically up and down the floors of a skyscraper. Indeed, it remains an open question how to optimize the research design in different geometries and to what extent the performance that we document here generalizes. 

It is our hope that the SFD research design reopens closed doors in the analysis of cross-sectional data. In many fields of economics---such as environment, development, geography, health, industrial organization, labor, public, growth, trade, and urban---there are core questions that are fundamentally cross-sectional in nature. Historically, econometricians that seek to address these questions have generally had two options: to trust that unobserved heterogeneity does not confound the analysis or to employ cross-sectional instruments which rely on exclusion restrictions that cannot be tested. The SFD research design may offer yet another path.

  % -----------------------------------------------
% 	FIGURES
% -----------------------------------------------

 % -----------------------------------------------
% 	REFERENCES
% -----------------------------------------------

%\newpage
\singlespacing
\bibliography{SFD_refs.bib}

%-------------------------------------APPENDIX
%\newpage
\singlespacing

\newpage
\appendix
\noindent {\bf {\LARGE Appendix}}
%\noindent {\bf {\LARGE Appendix}}
\vspace{3mm}

\renewcommand{\thesection}{A\arabic{section}} 
\setcounter{section}{0}
\setcounter{table}{0}
\setcounter{figure}{0}
\renewcommand\thefigure{A\arabic{figure}}
\renewcommand\thetable{A\arabic{table}}

 \section{Alternative interpretation of the identifying assumption}

It may not always be natural or intuitive to consider whether $\Delta \mb{x}$ and $\Delta \mb{c}$ are orthogonal in some settings.  In some of these cases, it may be more natural to rewrite the identifying assumption in Eq. (\ref{Eq:IndepCondition}) as
 \begin{align}
E[\underbrace{(\mb{x}'_i-\mb{x}'_{i-1})}_{\Delta\mb{x}'_i}\Delta\mb{c}_i]&=0_{K,M} \notag \\
E[\mb{x}_i'\Delta\mb{c}_i]&=E[\mb{x}_{i-1}'\Delta\mb{c}_i]
\label{Eq:Symmetry}
 \end{align}
which says that the first differences of any omitted variables are equally correlated with the levels of the regressor, regardless of whether one examines the observation at the ``beginning'' ($i-1$) or ``ending'' ($i$) of each pair of observations used to construct the SFD.  More intuitively, Eq. (\ref{Eq:Symmetry}) says that SFD will be identified if an observer standing at $i$ and observing $\mb{x}_i$ while looking ``West'' toward $i-1$ will have no more information about the change in omitted variables $\Delta\mb{c}_i$ between $i$ and $i-1$ than if she stood at $i-1$ and observed $\mb{x}_{i-1}$ while looking ``East.''  Checking Eq. (\ref{Eq:Symmetry}) may be natural in some cases, for example, in the 10th Avenue example, we think it is plausible that observing the average years of schooling in one census block ($\mb{x}$) provides no more information about the change in racial composition between census blocks ($\Delta \mb{c}$) when looking in the Uptown direction relative to looking in the Downtown direction. 

An expression analogous to Eq. (\ref{Eq:Symmetry}) using levels of $\mb{c}$ and the first differences of $\mb{x}$ can also be written, which may be helpful in some cases. 

\section{Comparison of SFD and Robinson's semi-parametric approach when estimating returns to schooling along 10th Ave and I-90\label{AppendixRobinson}}

We explore how the SFD estimator compares to the semi-parametric model  proposed by Robinson (1998) in the context of our returns to schooling example. Specifically, we estimate non-parametric ``spatial trends" in \textit{log wages} and \textit{years of education} across census tracts using kernel estimators, and then regress the resulting residuals of \textit{log wages} on the residuals of \textit{years of education}. We employ a uniform kernel with diminishing bandwidths ($h=3$, $h=2$, $h=1$). We expect the Robinson estimator to approach the SFD estimates as bandwidths become smaller, although the two should not be identical. Note that because we use a uniform kernel to implement the Robinson estimator, this approach is identical to using ``spatial fixed effects''  \citep{conley2010}.

The results are shown in Table \ref{Tab:Robinson}. The semi-elasticities estimated using Robinson's approach and a bandwidths of $h=3$ are 0.98 in Manhattan and 0.138 in Chicago. While the New York estimate is comparable to previous estimates of the return to education, the estimate in Chicago is larger than all 17 estimates of the return to education in the United States reported in Card (2001), which range from 0.052 to 0.132. When we instead use a bandwidth of $h=2$, the estimated effects decline slightly,  to 0.093 in New York and 0.110 in Chicago. With a bandwidth of $h=1$, the estimated effect in New York (0.081) is near the SFD estimate of 0.087; however, the estimated effect in Chicago falls to 0.042 and is estimated imprecisely. 

The difference between the SFD estimates and the estimates produced using Robinson's approach arise from different identifying assumptions. Under Robinson's approach, one must assume that all census tracts near enough to tract $i$ to inform the kernel estimates at location $\ell_i$ are comparable to $i$.  In New York and Chicago, wages and years of education are highly variable across space, which makes this assumption difficult to defend. Nonetheless, the similarity of these results, where SFD estimates are statistically indistinguishable from those using Robinson's method for a bandwidth of one, reinforces the interpretation of SFD as non-parametrically removing a highly flexible spatial trend from a partially linear model.

%\bigskip

\begin{table}[h!]
\thispagestyle{empty}
\fontsize{9}{9}\selectfont
\begin{center}
\newcolumntype{Y}{>{\centering\arraybackslash}X}
\begin{tabularx}{\linewidth}{ l Y Y Y Y Y Y Y Y}
\hline
\hline
  & & & & & & & & \\ 
& \multicolumn{8}{c}{\textit{Dependent variable: log average wage}} \\
  & & & & & & & & \\ 
\cline{2-9} 
  & & & & & & & & \\ 
& \multicolumn{4}{c}{10th Avenue, New York} & \multicolumn{4}{c}{I-90, Chicago} \\
  & & & & & & & & \\ 
& \multicolumn{3}{c}{Robinson} & SFD & \multicolumn{3}{c}{Robinson} & SFD \\
& ($h=3$) & ($h=2$) & ($h=1$) & & ($h=3$) & ($h=2$) & ($h=1$) & \\
  & & & & & & & & \\ 
 \hline
  & & & & & & & & \\ 
 Average years of eduction & 0.098$^{***}$ & 0.093$^{***}$ & 0.081$^{***}$ & 0.087$^{***}$ & 0.138$^{***}$ & 0.110$^{***}$ & 0.042 &  0.072$^{*}$  \\ 
  & (0.023) & (0.024) & (0.029) & (0.027) & (0.029) & (0.033) & (0.039) & (0.037) \\ 
  & & & & & & & & \\ 
 Constant & 0.002 & 0.003 & 0.002 & $-$0.010 & 0.003 & 0.002 & 0.0005 & $-$0.0002 \\ 
  & (0.025) & (0.024) & (0.019) & (0.039) & (0.023) & (0.021) & (0.016) & (0.035) \\ 
  & & & & & & & & \\ 
 \hline
  & & & & & & & & \\ 
Observations & 54 & 54 & 54 & 53 & 54 & 54 & 54 & 53 \\ 
R squared & 0.259 & 0.221 & 0.134 & 0.164 & 0.301 & 0.175 & 0.021 & 0.070 \\ 
  & & & & & & & & \\ 
 \hline
 \hline
 \end{tabularx}
 \caption{{\bf Cross-sectional estimates for returns to education Robinson and SFD.} Data are for census tracts in Manhattan, New York along 10th Avenue (columns 1-4) and Chicago, Illinois along Interstate-90 (columns 5-8) for the year 2010. Bandwidths $h$ are in units of census blocks. See text for details. Asterisks indicate statistical significance at the 0.1\%,***, 1\%**, and 5\%* levels.}
 \label{Tab:Robinson}
\end{center}
\end{table}

%\newpage
\section{Calculation of different standard error estimates for maize yields \label{AppendixSE}}

In our SFD estimation of the effect of climate and soil on maize yields, we report standard errors accounting for spatial autocorrelation following Conley (1999). Here, we explore alternative approaches to estimating the covariance matrix for SFD estimates. We expect the residual $\hat{\Delta \epsilon}$ to be negatively serial correlated, since the error terms of two sequential differenced observations, $\Delta \epsilon_i = \epsilon_i - \epsilon_{i-1}$ and $\Delta \epsilon_{i+1} = \epsilon_{i+1} - \epsilon_{i}$ both contain $\epsilon_i$ and this term enters positively in one instance and negatively in the other.  However, it is not clear \textit{ex ante} whether there exist correlations in $\hat{\Delta \epsilon}$ across larger distances, either among units within a channel or across channels.

To explore how the reported Conley standard errors compare to other common procedures for overcoming autocorrelation and heteroskedasticy in disturbances, we calculate four different sets of standard errors: (i) Newey-West standard errors, (ii) clustered standard errors (iii) bootstrapped standard errors, and (iv) block bootstrapped standard errors. The clustered standard errors are clustered by sampling channel. For the bootstrapped standard errors, we resample at the observation-level after differencing between adjacent counties. For the block bootstrap, we resample entire sequences of differenced observations for each channel. These standard errors are presented in Table \ref{Tab:StandardErrors}, along with the OLS standard errors for comparison. The clustered and bootstrapped standard errors are comparable in magnitude to the OLS standard errors, while the Newey-West and block bootstrapped standard errors tend to be larger. 

Note that Newey-West and Conley approaches are identical in one-dimensional spaces, but differ in two dimensional spaces. This is because the Newey-West approach restricts autocorrelation to be estimated only along a sampling channel, whereas the Conley approach allows for autocorrelation among $\hat{\Delta \epsilon}$ that are near one another in physical space but not contained within the same channel sequence. The block boostrapping approach accounts for within-channel autocorrelation, similar to Newey-West, but not across-channel correlations. In the main text, we report Conley standard errors because they appear to be the largest and most conservative in general, suggesting there may be some cross-channel autocorrelation in $\hat{\Delta \epsilon}$ in the context of US maize yields. 

\begin{table}
\vspace{-2cm}
\fontsize{8}{8}\selectfont
\begin{center}
\begin{tabular}{ l c c c c c c c c }

\hline
\hline
 & & & & & & & &  \\
& \multicolumn{8}{c}{\textit{Dependent variable: average log maize yield $\times$ 1,000}} \\
 & & & & & & & & \\
\cline{2-9} 
 & & & & & & & &  \\
 & (1 WE) & (1 NS) & (2 WE) & (2 NS) & (3 WE) & (3 NS) & (7 WE) & (7 NS) \\
 & & & & & & & &  \\
 \hline
 & & & & & & & &  \\
 
 Degree days (below 29\degree C) & 0.35 & 0.43 &  &  &  &  & 0.28 & 0.33 \\ 
  & (0.12) & (0.11) &  &  &  &  & (0.10) & (0.10) \\ 
& $\langle 0.08 \rangle$ & $\langle 0.08 \rangle$ & & & & & $\langle 0.07 \rangle$ & $\langle 0.07 \rangle$ \\ 
& [0.09] & [0.09] & & & & & [0.08] & [0.08] \\ 
& \{0.12\} & \{0.13\} & & & & & \{0.11\} &  \{0.12\} \\ 
& ((0.08)) & ((0.08)) & & & & & ((0.07)) & ((0.07)) \\ 
 & & & & & & & &  \\
 
 Degree days (above 29\degree C)  & $-$4.99 & $-$4.77 &  &  &  &  & $-$4.73 & $-$4.41 \\ 
  & (1.57) & (1.48) &  &  &  &  & (1.33) & (1.33)  \\ 
& $\langle 0.94 \rangle$ & $\langle 0.92 \rangle$ & & & & & $\langle 0.86 \rangle$ & $\langle 0.87 \rangle$ \\ 
& [1.19] & [1.17] & & & & & [1.14] & [1.11] \\ 
& \{1.63\} & \{1.78\} & & & & & \{1.57\} &  \{1.55\} \\ 
& ((0.92)) & ((0.91)) & & & & & ((0.84)) & ((0.85)) \\ 
 & & & & & & & &  \\
 
 Precipitation &  &  & 3.72 & 4.10 &  &  & 2.52 & 2.99 \\ 
  &  &  & (1.23) & (1.06) &  &  & (1.11) & (0.90) \\ 
& & & $\langle 1.15 \rangle$ & $\langle 0.88 \rangle$& & & $\langle 1.01 \rangle$ & $\langle 0.79 \rangle$ \\ 
& & & [1.14] & [1.27] & & & [1.08] & [1.16] \\ 
& & & \{1.51\} & \{1.80\} & & & \{1.39\} &  \{1.45\} \\ 
& & & ((1.13)) & ((1.08)) & & & ((0.87)) & ((0.78))  \\ 
 & & & & & & & &  \\
 
 Precipitation squared &  &  & $-$0.003 & $-$0.003 &  &  & $-$0.002 & $-$0.002 \\ 
  &  &  & (0.001) & (0.001) &  &  & (0.001) & (0.001) \\ 
& & & $\langle 0.001 \rangle$ & $\langle 0.0007 \rangle$& & & $\langle 0.0008 \rangle$ & $\langle 0.0006 \rangle$ \\ 
& & & [0.001] & [0.001] & & & [0.001] & [0.001] \\ 
& & & \{0.001\} & \{0.001\} & & & \{0.001\} &  \{0.001\} \\ 
& & & ((0.001)) & ((0.0007)) & & & ((0.0008)) & ((0.0006))  \\ 
 & & & & & & & &  \\
 
 Water capacity &  &  &  &  & 19.82 & 19.26 & 18.37 & 17.09 \\ 
  &  &  &  &  & (3.61) & (4.10) & (3.63) & (3.84) \\ 
& & & & & $\langle 2.72 \rangle$ & $\langle 2.64 \rangle$&$\langle 2.69 \rangle$ & $\langle 2.61 \rangle$ \\ 
& & & & & [3.27] & [3.02] & [3.18] & [2.96] \\ 
& & & & & \{3.70\} & \{4.48\} & \{3.95\} &  \{4.22\} \\ 
& & & & & ((2.66)) & ((2.60)) & ((2.63)) & ((2.56)) \\ 
 & & & & & & & &  \\
 
 Percent clay &  &  &  &  & $-$0.43 & $-$2.01 & $-$0.13 & $-$1.67 \\ 
  &  &  &  &  & (1.01) & (0.96) & (0.98) & (0.98) \\ 
& & & & & $\langle 0.74 \rangle$ & $\langle 0.73 \rangle$&$\langle 0.73 \rangle$ & $\langle 0.71 \rangle$ \\ 
& & & & & [0.88] & [0.86] & [0.85] & [0.84] \\ 
& & & & & \{0.98\} & \{1.01\} & \{0.87\} &  \{0.97\} \\ 
& & & & & ((0.72)) & ((0.71)) & ((0.71)) & ((0.70)) \\ 
 & & & & & & & &  \\
 
 Minimum permeability &  &  &  &  & 16.67 & 14.01 & 17.65 & 15.00 \\ 
  &  &  &  &  & (6.33) & (7.20) & (5.84) & (6.73) \\ 
& & & & & $\langle 4.20 \rangle$ & $\langle 4.19 \rangle$&$\langle 4.10 \rangle$ & $\langle 4.08 \rangle$ \\ 
& & & & & [5.39] & [6.75] & [4.97] & [6.31] \\ 
& & & & & \{7.20\} & \{7.74\} & \{6.67\} &  \{7.19\} \\ 
& & & & & ((4.11)) & ((4.12)) & ((4.01)) & ((4.00)) \\ 
 & & & & & & & &  \\
 
 Soil erodibility factor &  &  &  &  & $-$348.43 & $-$286.06 & $-$335.54 & $-$278.18 \\ 
  &  &  &  &  &  (192.72) & (165.01) & (182.96) & (159.39) \\ 
& & & & & $\langle 105.8 \rangle$ & $\langle 109.0 \rangle$&$\langle 103.5 \rangle$ & $\langle 106.4 \rangle$ \\ 
& & & & & [154.3] & [128.1] & [145.9] & [127.1] \\ 
& & & & & \{209.6\} & \{192.4\} & \{198.1\} &  \{192.6\} \\ 
& & & & & ((103.6)) & ((107.2)) & ((101.2)) & ((104.3)) \\ 
 & & & & & & & &  \\
 
 Best soil class &  &  &  &  & 2.32 & 2.19 & 2.43 & 2.29 \\ 
  &  &  &  &  & (0.36) & (0.54) & (0.32) & (0.44) \\ 
& & & & & $\langle 0.24 \rangle$ & $\langle 0.23 \rangle$&$\langle 0.24 \rangle$ & $\langle 0.23 \rangle$ \\ 
& & & & & [0.33] & [0.33] & [0.29] & [0.29] \\ 
& & & & & \{0.34\} & \{0.58\} & \{0.32\} &  \{0.46\} \\ 
& & & & & ((0.24)) & ((0.23)) & ((0.29)) & ((0.23)) \\ 
 & & & & & & & &  \\
 
 Constant & 4.99 & $-$4.50 & 6.45 & $-$3.47 & 9.75 & 2.41 & 5.61 & $-$3.46 \\ 
  & (2.44) & (2.77) & (2.19) & (2.56) & (2.04) & (2.17) & (1.91) & (2.97) \\ 
&  $\langle 3.02 \rangle$ & $\langle 3.32 \rangle$ &  $\langle 3.02 \rangle$ & $\langle 3.04 \rangle$ &  $\langle 2.64 \rangle$ & $\langle 2.64 \rangle$ &  $\langle 2.64 \rangle$ & $\langle 3.08 \rangle$ \\ 
& [3.01] & [3.30] & [2.92] & [2.89]  & [2.59] & [2.68]  & [2.56] & [3.10]  \\ 
& \{2.69\} & \{1.93\} & \{2.22\} & \{2.37\} & \{2.53\} & \{2.14\}  & \{2.23\} & \{2.68\}  \\ 
& ((2.97)) & ((3.27)) & ((2.96)) & ((2.97)) & ((2.58)) & ((2.59)) & ((2.59)) & ((3.02)) \\ 
 & & & & & & & &  \\
 
 \hline
 & & & & & & & &  \\
 Observations & 804 & 825 & 804 & 825 & 804 & 825 & 804 & 825  \\ 
 & & & & & & & &  \\
 \hline
 \hline
 \end{tabular}
 \caption{\textbf{Standard Errors for SFD Estimates.} Standard errors are calculated using five different methods: (Newey-West standard errors), $\langle$Clustered standard errors$\rangle$, [Bootstrapped standard errors], \{Block bootstrapped standard errors\}, and ((OLS standard errors)). This Table corresponds to Table \ref{Tab:SFD} in the main text.}
\label{Tab:StandardErrors}
\end{center}
\end{table}

\end{document}